\documentclass{pasj00}
\SetRunningHead{A Novel Method to Identify AGNs}{Tanaka}

\usepackage{times}
\usepackage{ulem}

\begin{document}
\title{A Novel Method to Identify AGNs Based on Emission Line Excess and\\the Nature of Low-luminosity AGNs in the Sloan Digital Sky Survey\\I -- A Novel Method}
\author{Masayuki Tanaka}
\affil{Institute for the Physics and Mathematics of the Universe, The University of Tokyo\\  5-1-5 Kashiwanoha, Kashiwa-shi, Chiba 277-8583, Japan}
\email{masayuki.tanaka@ipmu.jp}
\KeyWords{galaxies: active --- methods: data analysis --- X-rays: galaxies --- radio continuum: galaxies}

\maketitle

\begin{abstract}
We develop a novel technique to identify active galactic nuclei (AGNs) and
study the nature of low-luminosity AGNs in the Sloan Digital Sky Survey.
This is the first part of a series of papers and we develop a new, sensitive
method to identify  AGNs in this paper.
An emission line luminosity in a spectrum is a sum of a star formation component
and an AGN component (if present).
We demonstrate that an accurate estimate of the star formation component can be
achieved by fitting model spectra, generated with a recent stellar population synthesis code,
to a continuum spectrum.
By comparing the observed total line luminosity with that attributed to star formation,
we can tell whether a galaxy host an AGN or not.
We compare our method with the commonly used emission line diagnostics proposed by
Baldwin et al. (1981; hereafter BPT).
Our method recovers the same star formation/AGN classification
as BPT for 85\% of the strong emission line objects, which comprise 43\% of our sample.
A unique feature of our method is its sensitivity: it is applicable to 78\% of the sample.
We further make comparisons between our method and BPT
using stacked spectra and selection in X-ray and radio wavelengths.
We show that, while the method suffers from incompleteness and contamination as
any AGN identification methods do, it is overall a sensitive method
to identify AGNs.
We emphasize that the method can be applied at high redshifts (up to $z\sim1.7$ with
red-sensitive optical spectrograph) without making any a priori assumptions about
host galaxy properties.
Another unique feature of the method is that it allows us to subtract emission line
luminosity due to star formation and extract intrinsic AGN luminosity.
We will make a full use of these features to study the nature of low-luminosity
AGNs in Paper-II.
\end{abstract}

\section{Introduction}

Since the seminal work by \citet{seyfert43} on broad emission line
properties of spiral nebulae, there has been a rapid
advance in our understanding of active galactic nuclei (AGNs).
A number of spectroscopic surveys of nuclei of nearby galaxies
have been performed since then (e.g., \cite{heckman80,ho95}) and 
large surveys of the local universe such as the 2 degree
field survey \citep{colless03} and the Sloan Digital Sky Survey (SDSS; 
\cite{york00}) have characterized properties of AGNs with an unprecedented
statistical accuracy (e.g., \cite{kauffmann03,heckman04,kewley06,schawinski07}).
The realization that black hole mass correlates with host galaxy
properties such as bulge luminosity and mass
\citep{kormendy95,magorrian98,ferrarese00,gebhardt00} has triggered
a lot of effort to link super-massive black hole growth with galaxy growth.
Recent simulations of galaxy evolution seem to achieve some success
in reproducing observed properties of galaxies by incorporating
energy feedback from AGNs, although details of such a feedback
mechanism are still fairly uncertain \citep{granato04,croton06,bower06}.

In order to perform an observational 
study of effects of AGNs on galaxy evolution,
one first has to identify AGNs.
AGNs can be identified in a variety of ways at essentially all wavelengths
(X-rays, mid-IR, and radio detections and optical emission line
diagnostics).
The energy source of AGN activity is the material accreting
onto a central super-massive black hole that is
heated to high temperature, traveling at high velocity.
It is thought to be the source of X-ray emission due to
the inverse-Compton scattering of thermal photons from the accretion disk
off relativistic electrons.
The surrounding dust torus is heated by the radiation, thus
resulting in strong thermal emission at mid-IR wavelengths.
In some cases, the central engine ejects jets perpendicular to the accretion
disk and they are often observed in radio wavelengths due
to the Synchrotron radiation.
Also, the central engine ionizes the gas in the galaxy
out to hundred-parsec to kilo-parsec scales.
The ionization state of the surrounding gas is
often higher than normal star forming regions, showing
characteristic emission line intensity ratios.
AGNs exhibit these unique spectral features and AGN identification
methods aim at detecting them.

In this paper, we focus on optical emission line techniques.
\citet{baldwin81} first presented a method to separate
AGNs from star forming galaxies using flux ratios of
four emission lines in the optical wavelengths
(H$\beta$, {\sc [oiii]}, H$\alpha$, and {\sc [nii]}).
We refer to this diagnostics as BPT in what follows.
AGNs and star forming galaxies form distinct sequences
with some overlap on the BPT diagram.
Theoretical calculations based on photo-ionization models
have been made to understand
the distribution on this diagram (e.g., \cite{ho93,kewley01,stasinska06}).
For instance, \citet{kewley01} performed detailed photo-ionization modeling
over a range of ionization parameters and defined a region
of the diagram where photo-ionization is unlikely due
to stars.
Massive spectroscopic surveys of the local universe
such as the Sloan Digital Sky Survey (SDSS; \cite{york00})
revealed that star forming galaxies form a tight
sequence on the BPT diagram, suggesting that the ionization
state of star forming galaxies does not vary so widely
as explored by \citet{kewley01}.
\citet{kauffmann03} empirically revised the theoretical
curve of \citet{kewley01} using the data from SDSS and
this is now a commonly used discriminator between
AGNs and star forming galaxies.

Despite the popularity, however, the BPT diagnostics has
disadvantages.  Firstly, it requires four emission lines,
which are not always easy to measure with sufficient
signal-to-noise ratios. 
In particular, H$\beta$ can be very weak in low-luminosity AGNs
and it often limits the sensitivity of BPT.
Secondly, it requires H$\alpha$ and {\sc [nii]}.
These lines migrate to near-IR at $z>0.5$ and
it is hard to measure these lines at high redshifts.
Attempts have been made to overcome these issues
by using emission lines that are observable even at $z>0.5$
or by making a priori assumption of host galaxy properties.
\citet{rola97} suggested that {\sc [oii]} and {\sc [neiii]}$\lambda\lambda 3869,3968$
could help identify AGNs when H$\alpha$ and {\sc [nii]} are not available.
\citet{lamareille04} showed that {\sc [oii]}/H$\beta$ can
be used in place of {\sc [nii]}/H$\alpha$.
\citet{yan06} and \citet{yan11} proposed ways to identify
AGNs using a priori assumption about host galaxy properties.
Recently, \citet{juneau11} presented a new diagnostics using {\sc [oiii]}/H$\beta$
and stellar mass of the hosts.
In this paper, we make an attempt to overcome the issues
with a new, physically motivated method to identity AGNs.
In particular, we do not assume any host galaxy properties
a priori as done in Yan et al. (2006, 2011) and \citet{juneau11}
to identify AGNs.
This is essential to study relationships between the AGN activity
and host galaxy properties.

The structure of this paper is as follows.
We develop a new method to identify AGNs in Section 2, followed
by extensive tests of the method in Section 3.
We summarize the strengths and weaknesses of the method and
conclude the paper in Section 4.  Nature of low-luminosity
AGNs and their host galaxy properties will be presented in Paper-II.
Unless otherwise stated, we adopt $\Omega_M=0.3$,
$\Omega_\Lambda=0.7$, and $\rm H_0=70\ km\ s^{-1}\ Mpc^{-1}$.
All the magnitudes are given in the AB system.
We use the following abbreviations : AGN for active galactic
nucleus, BPT for the \citet{baldwin81} diagnostics,
SF for star formation, and SFR for star formation rate.
Emission lines used in this work include {\sc [oii] $\lambda\lambda3726,3729$},
H$\beta\ \lambda4861$, {\sc [oiii] $\lambda5007$}, {\sc [oi] $\lambda6300$},
H$\alpha\ \lambda6563$, {\sc [nii] $\lambda6583$}, and {\sc [sii] $\lambda6716,6730$}.

\section{A new method}

As mentioned in the last section, the commonly used emission line
diagnostics involves intensity ratios of emission lines to identify a signature of AGN.
An ionizing spectrum of AGN is typically harder than
spectra of young stars, and thus AGNs exhibit characteristic emission
line intensity ratios.  The most commonly adopted \citet{baldwin81}
diagnostics involves four emission lines, which are not always
easy to measure at high signal-to-noise.  That hinders efficient
identification of low-luminosity AGNs in surveys such as SDSS.
However, one does not necessarily have to rely on ratios of emission lines.
A single emission line in principle contains information about an underlying AGN.
If a galaxy hosts an AGN, the emission
line luminosity we observe originates both from star formation and AGN:

\begin{equation}
L_{measured} = L_{SF} + L_{AGN},
\end{equation}

\noindent
where $L_{SF}$ is an emission line luminosity due to star formation
and $L_{AGN}$ is a luminosity due to AGN.  
The idea behind our method is to estimate $L_{SF}$ of a galaxy and compare
it with $L_{measured}$.  If we observe a significant luminosity excess
in $L_{measured}$, it means that the galaxy shows a significant $L_{AGN}$
and it likely hosts an AGN.
In this section, we develop a method to estimate $L_{SF}$
and quantify how accurate our $L_{SF}$ is.
Then we move on to perform an extensive test of
our AGN identification method in the next section.

\subsection{Sloan Digital Sky Survey}

In this paper, we use data from the Sloan Digital Sky Survey
Data Release 7 \citep{abazajian09}.
The SDSS utilizes a dedicated 2.5m telescope installed
at the Apache Point Observatory \citep{gunn06} and the survey is
in two parts: imaging and spectroscopy.
The SDSS has imaged a quarter of the sky in five photometric bands
($urgiz$; \cite{fukugita96,gunn98,doi10}) with unprecedented
accuracy \citep{ivezic07,padmanabhan08}.
The SDSS spectroscopic survey utilizes double fiber-fed
spectrographs and obtains 640 spectra simultaneously covering
a wavelength range of 3800\AA\  to 9200\AA\  with a resolving
power of $R\sim2000$.
Each fiber subtends $3''$ on the sky.   
The survey consists of 3 major components : 
main galaxy sample \citep{strauss02}, luminous red galaxy sample
\citep{eisenstein01}, and QSO sample \citep{richards02}.
The main sample is a flux-limited sample down to $r=17.77$ selected from
the imaging survey and we use objects in the main sample in this paper.

We apply the following criteria to select galaxies for our study:
{\sc specClass=2} (i.e., objects are galaxies) located at
$0.02<z<0.10$  with high confidence flags ({\sc zConf$>$0.8}
and {\sc zWarning=0}).
We intentionally remove QSO-like objects ({\sc specClass=3})
from the sample  because our method is
not applicable to those objects (as shown below, our method assumes
that a continuum spectrum is dominated by stars, not by AGN).
We have 283,031 objects in total, a quarter of which are
identified as AGNs by the method developed in this paper.
We correct the SDSS spectra for the Galactic extinction using
the extinction curve of \citet{cardelli89} and the extinction
map from \citet{schlegel98}.

\subsection{Spectral fitting}

How do we estimate $L_{SF}$ of an emission line?
It has actually been a long standing issue in AGN studies.
AGN emission is contaminated with star formation emission and that has
often hindered detailed studies of intrinsic AGN output.
We cannot use any emission lines to estimate $L_{SF}$ because
it is hard to discriminate it from $L_{AGN}$.
We have to rely on other available information.
One might use an star formation indicator such as
mid-IR emission to measure it, but one faces the same problem --
it is difficult to disentangle $L_{SF}$ and $L_{AGN}$ in mid-IR.
We take a novel approach to solve the problem.
AGNs show strong emission lines, but their continuum emission is
usually very weak in optical wavelengths, except for very strong
AGNs and quasars \citep{binette94}.
We use optical continuum emission of galaxies, which is dominated
by stellar light, not by AGNs, to estimate $L_{SF}$.
We fit observed spectra of galaxies with model spectral templates of
galaxies with various star formation histories to obtain star formation
rates (SFRs) and dust extinction.  From these numbers, we can work out
$L_{SF}$.

We generate model templates using an updated version of the 
\citet{bruzual03} code with improved treatment of thermally pulsating
AGB stars.  Free parameters in the models are

\begin{itemize}
\item {\bf Star formation history:}
We assume a simple, exponentially decaying star formation rate
to describe star formation histories.
The exponential time scale is allowed to vary between 0 
(i.e., instantaneous burst) and $\infty$ (i.e., constant star formation rate).

\item {\bf Dust extinction:}
We use the two component extinction model of \citet{charlot00}.
We adopt $\mu=0.3$, which means that 30\% of the extinction is
due to the ambient interstellar medium, which affects all stars.
The remaining 70\% is due to dust in star forming regions,
and it affects only stars younger than $10^7$ yr.
We modify the extinction curve of $\tau\propto\lambda^{-0.7}$
to that of \citet{cardelli89}, which is close to
$\tau\propto\lambda^{-1.0}$.
We justify the choice of the \citet{cardelli89}
curve over the \citet{charlot00} curve in the Appendix.
We allow the optical depth in the $V$-band, $\tau_V$, to
vary between 0 (i.e., no dust) and 3.

\item {\bf Metallicity:}
We use the solar metallicity models only.  If we include
super-solar and sub-solar metallicity models, we introduce
too much degeneracies between age, metallicity, and dust
and degrade the fits.
We justify the exclusion of non-solar metallicity models
in the Appendix.

\item {\bf Age:} We apply a logical constraint that
the age of an model template must be younger than the age
of the universe at a given redshift.  We do not use models
with young ages with $<1$ Gyr.
Due to the degeneracies between the above mentioned parameters,
we can fit galaxies with very young models, but they often give
inaccurate SFRs.  The age limit of 1 Gyr removes most of
such bad fits.
\end{itemize}

\noindent
We assume the Chabrier initial mass function \citep{chabrier03}.
We fit the observed spectra of galaxies with these templates
using the $\chi^2$ statistics.
We generate sets of the templates with varying stellar velocity
dispersions ranging from 75 to 250 $\rm km\ s^{-1}$ with
a $25\rm km\ s^{-1}$ step and fit the galaxies with the closest dispersion.
We have $\sim 8,400$ model templates in each set.
In the fitting, we mask out regions around strong emission lines
such as {\sc [oii]}, H$\beta$, {\sc [oiii]}, H$\alpha$,
{\sc [nii]}, {\sc [sii]}, etc, because we want to fit the spectra of stars.
In addition, we mask out a region around 5577$\rm\AA$, where a strong
night-sky Oxygen line is located.
The best-fitting models give SFRs, stellar mass, and $\tau_V$,
which we will extensively use in our analysis.
We derive an uncertainty on each parameter by taking $\Delta\chi^2=1$
from the best fit.  However, due to correlations between adjacent
wavelength points of the SDSS spectra and also to strong degeneracies
between the model parameters, the derived errors may not be
accurate.  We can empirically measure the errors in, e.g.,
$L_{SF}$ by comparing those from direct emission line measurements
as shown below.

We subtract the best-fitting model spectra from the observed
spectra to obtain continuum-subtracted spectra to measure emission line fluxes.
As summarized by \citet{tojeiro10},
population synthesis models are not always perfect and they under/over subtract
the continuum in some wavelength regions.
For example, in the case of \citet{bruzual03} model, it often over-subtracts
the continuum around H$\beta$ \citep{asari07}.
We therefore remove the residual continuum by median-filtering
the spectra within a running box of $\Delta\lambda=\rm60\AA$.
We then simply sum the fluxes around an emission line to measure
its flux within a wavelength range of $|\lambda-\lambda_{line}|<8\rm\AA$.
One could fit Gaussian to an emission line to measure the flux (e.g.,
\cite{tremonti04}), but AGNs may well show narrow and broad components
simultaneously.  As AGNs are the focus of this work, we do not assume any line profiles.
A fraction of galaxy light is missed from the 3 arcsec fiber.
In order to correct for the missing light, we compute the slit loss
by comparing $r$-band magnitude synthesized
from the spectrum with the $r$-band Petrosian magnitude from imaging.
The stellar masses and SFRs of the host galaxies are corrected for
the slit losses and are indicated with a subscript {\it apercorr} in figures.
Note that this is only a first-order correction because of the assumption
employed here that the light in the fiber is representative of the entire galaxy light.

Strong AGNs often exhibit featureless continuum in the UV (e.g., \cite{kinney91}),
which might affect our spectral fitting and the resultant parameters.
We have checked effects of such featureless continuum by including additional
continuum flux in the form of $F_\nu\propto\nu^{-1.5}$ \citep{schmitt99} to
the stellar spectra generated with the population synthesis code.
The strength of the featureless continuum ranges from 0 to 30\% of an observed spectrum
at 5500\AA\ with a step of 5\%.  We find that such a continuum decreases the accuracy
of our spectral fits.  For example, an accuracy of the predicted {\sc [oii]+[oiii]} 
luminosity
decreases to 0.35 dex (we obtain 0.24 dex without the continuum as shown below).
Furthermore, the best-fitting models often give strong featureless continuum
to star forming galaxies selected from BPT (i.e., non-AGNs).  These results suggest that such a continuum
just increases the degeneracies between the model parameters and does not
improve the fits.  It probably makes sense to include the featureless
continuum in the fits to study strong AGNs, but for our purpose of studying
low-luminosity AGNs, we choose not to include it.
The so-called 'big blue bump' seems to disappear in low-luminosity AGNs \citep{eracleous10a}.
This observation adds further motivation not to include the featureless continuum
in the fits.

We admit that there is room for improvements in our spectral fitting.
First, we do not use any priors in the fitting.
We may obtain better fits if we use priors on correlations between parameters
(e.g., one can assume a broad
correlation between dust, SFRs, and stellar mass), although it is not very straightforward
to set priors at high redshifts, where we actually would like
to apply the method in the future.
Also, one can model more realistic star formation histories by
including a secondary burst.
Despite the simplicity, however, our models deliver good estimates
of SFRs and emission line luminosities as shown below.

\subsection{Accuracy of SFRs and extinction from the spectral fits and the predicted $L_{SF}$}

\begin{figure*}
  \begin{center}
    \FigureFile(70mm,80mm){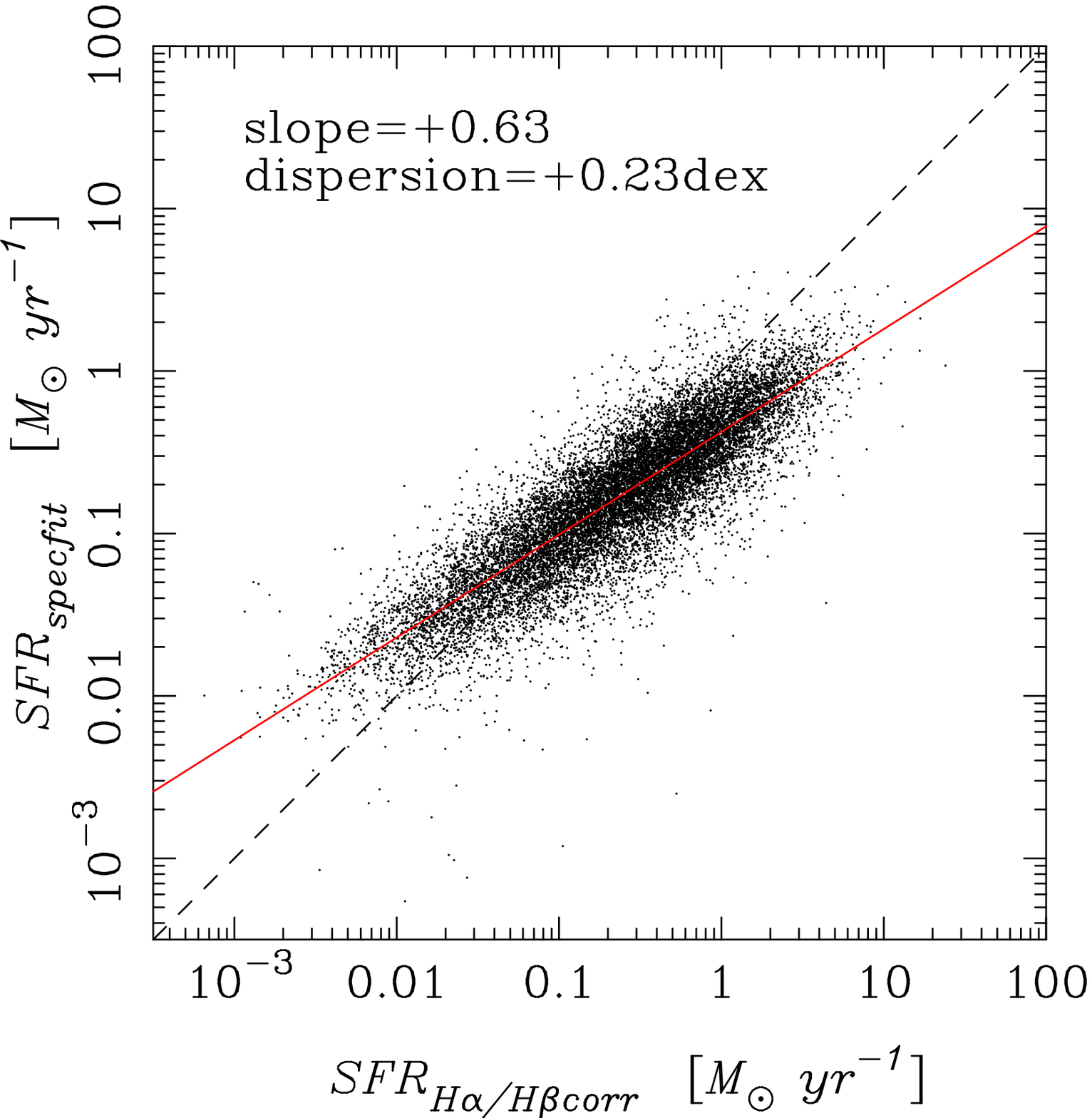}\hspace{0.5cm}
    \FigureFile(70mm,80mm){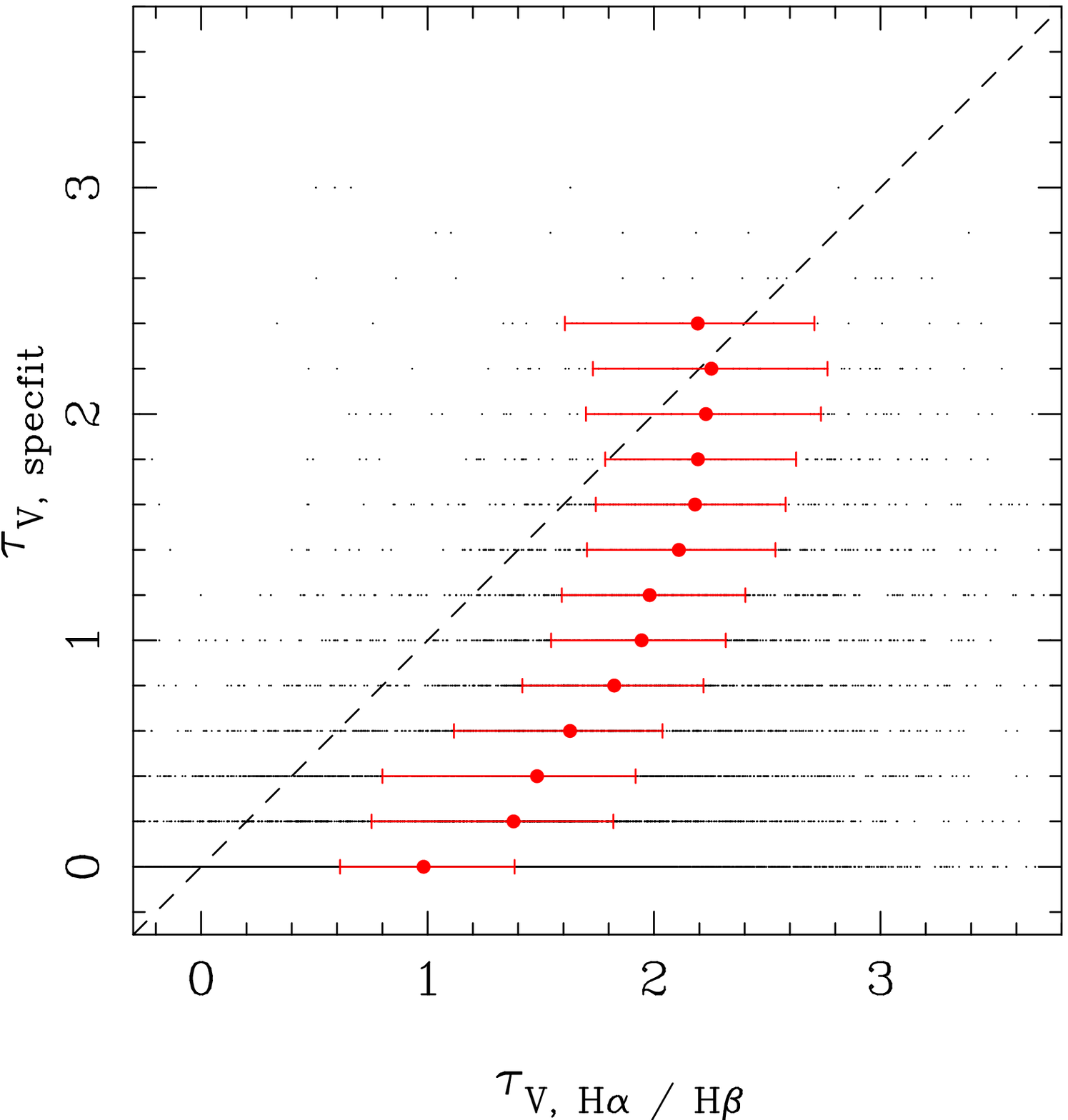}
  \end{center}
  \caption{
    {\bf Left:}
    SFRs from spectral fits plotted against SFRs from H$\alpha$
    corrected for the dust extinction using the balmer decrement
    assuming the extinction law of \citet{cardelli89} and intrinsic H$\alpha$/H$\beta$
    ratio of 2.86 \citep{osterbrock06}.  We plot every 5 objects
    for clarity.  The solid line shows a log-linear fit to the data
    and the slope and dispersion around the fit are shown in the plot.
    Note that AGNs identified with BPT are removed from the plot.
    {\bf Right:}
    $\tau_V$ from the spectral fits plotted against $\tau_V$
    from the balmer decrement.  The points and error bars show
    the median and 25th-75th percentile interval (they are shown
    only at $\tau_{V,specfit}$ bins with more than 100 galaxies).
  }
  \label{fig:model_vs_obs}
\end{figure*}

Using the SFRs and dust extinction measurements derived from
the model fits, we can now estimate $L_{SF}$.  But, first of all,
we shall quantify how accurate our estimate of SFRs and $\tau_V$ from
the model fits are.  In Fig. \ref{fig:model_vs_obs}, we compare
SFRs from the spectral fits and those from H$\alpha$ corrected
for extinction using the balmer decrement.
For the purpose of quantifying the accuracy of $L_{SF}$,
we remove AGNs using the BPT diagram with
the \citet{kauffmann03} threshold from the figure for now.
We use galaxies with all the H$\beta$, {\sc [oiii]},
H$\alpha$ and {\sc [nii]} lines detected at $>3\sigma$ here.

We obtain reasonably good estimates of SFRs from the spectral fits,
although there is a tilt there.
We apply a biweight fit and find a log-linear slope of 0.63 and
the scatter around the solid line to be only 0.23 dex (a factor of 1.7).
As discussed in the appendix, the tilt and the mean offset becomes slightly larger
if we use the original $\tau\propto\lambda^{-0.7}$ law of \citet{charlot00}.
Note as well that we obtain a better correlation with a log-linear slope
of 0.8 if we do not correct for the dust using the Balmer decrement.
We will extensively use SFRs from the spectral fits, and we could
empirically correct for the observed tilt to obtain more precise SFRs.
But, we choose not to do so because
it does not change our conclusions in the paper.
We try not to use empirical calibrations of 
the outputs from the spectral fitting throughout the paper as 
long as they do not affect our main conclusions.

While we can estimate SFRs reasonably well, our dust estimates are not
as good as we hoped for.
The two component dust model of \citet{charlot00} does not
seem to work very well and our dust measurements are
almost always smaller than those from the Balmer decrements
particularly at low $\tau_V$.
The median difference between the two extinction estimates
is $\tau_{v,\ specfit}-\tau_{v,\ H\alpha/H\beta}=-0.98$.

From SFR and $\tau_V$, we can work out $L_{SF}$.
In case of H$\alpha$, we use the following equation to derive it:

\begin{equation}
L_{H\alpha,\ SF}=SFR/(7.9\times10^{-42}/1.7)\times\exp(-0.75\tau_V)
\end{equation}

\begin{figure}
  \begin{center}
    \FigureFile(70mm,80mm){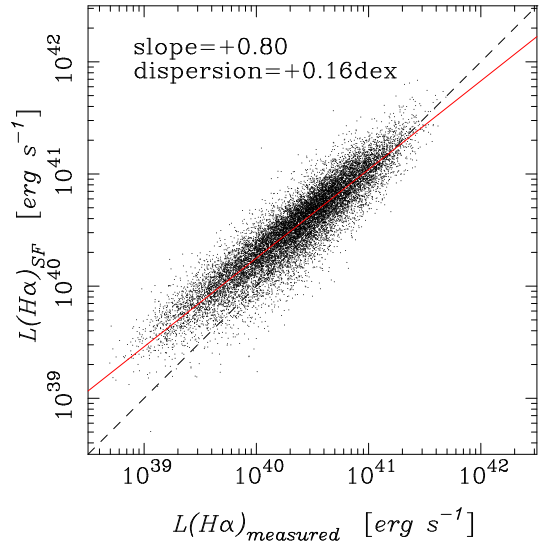}
  \end{center}
  \caption{
    Predicted H$\alpha$ luminosity from the spectral fits plotted against
    measured H$\alpha$ luminosity.  The slope of the log-linear fit and dispersion
    around it are shown in the plot.
  }
  \label{fig:ha_comp}
\end{figure}

\noindent
The conversion factor from SFR to H$\alpha$ flux is from \citet{kennicutt98}
and the factor of 1.7 is applied to change the initial mass function
from Salpeter (which is assumed in \cite{kennicutt98}) to Chabrier \citep{asari07}.
Extinction at the wavelength of H$\alpha$ is $0.75\tau_V$.
Fig. \ref{fig:ha_comp} compares the predicted H$\alpha$ luminosity
from the spectral fits with the measured luminosity.
The figure shows  that we can make a fairly accurate (a dispersion of 0.16 dex
or a factor of 1.4)
prediction of H$\alpha$ luminosity from the continuum fit.
This good correlation might appear surprising given the poor dust estimates,
but it is due to the degeneracies between SFR and dust --- slightly
underestimated SFRs and underestimated dust extinction nearly cancel out
and give us good emission line luminosity estimates.
We perform a biweight fit to the data and obtain a log-linear slope of
0.80.  This is a relatively small tilt and
and the accuracy of the predicted luminosity is encouraging.

We emphasize that Fig. \ref{fig:ha_comp} includes all the effects
from observations and model fits.
As discussed above, our models may not be the most
realistic models and there is room to improve the fitting procedure,
but we can still make a fairly good prediction  of $L_{SF}$.
Continuum may provide SFRs on different time scales
from those from H$\alpha$.  H$\alpha$ comes from star forming
regions and is a good probe of instantaneous SFRs, while continuum
may provide SFRs smoothed over an extended period.
The scatter we observe in Fig. \ref{fig:ha_comp} is perhaps partly
due to the different time scales probed.
But, including all these effects, we obtain a remarkable accuracy 
with only a small tilt.
We find that this scatter does not significantly reduce if we use
H$\alpha$ measured at $>10\sigma$ only, suggesting that the scatter
is primarily due to scatter in $L_{H\alpha, SF}$.
The errors in the physical parameters from the spectral fits may not be
accurate due to correlations between the adjacent wavelength points and
to model degeneracies as mentioned above, but the observed
dispersion gives us a good quantitative estimate of an error in
$L_{H\alpha, SF}$, which is a factor of 1.4 with a small tilt.

\begin{figure*}
  \begin{center}
    \FigureFile(80mm,80mm){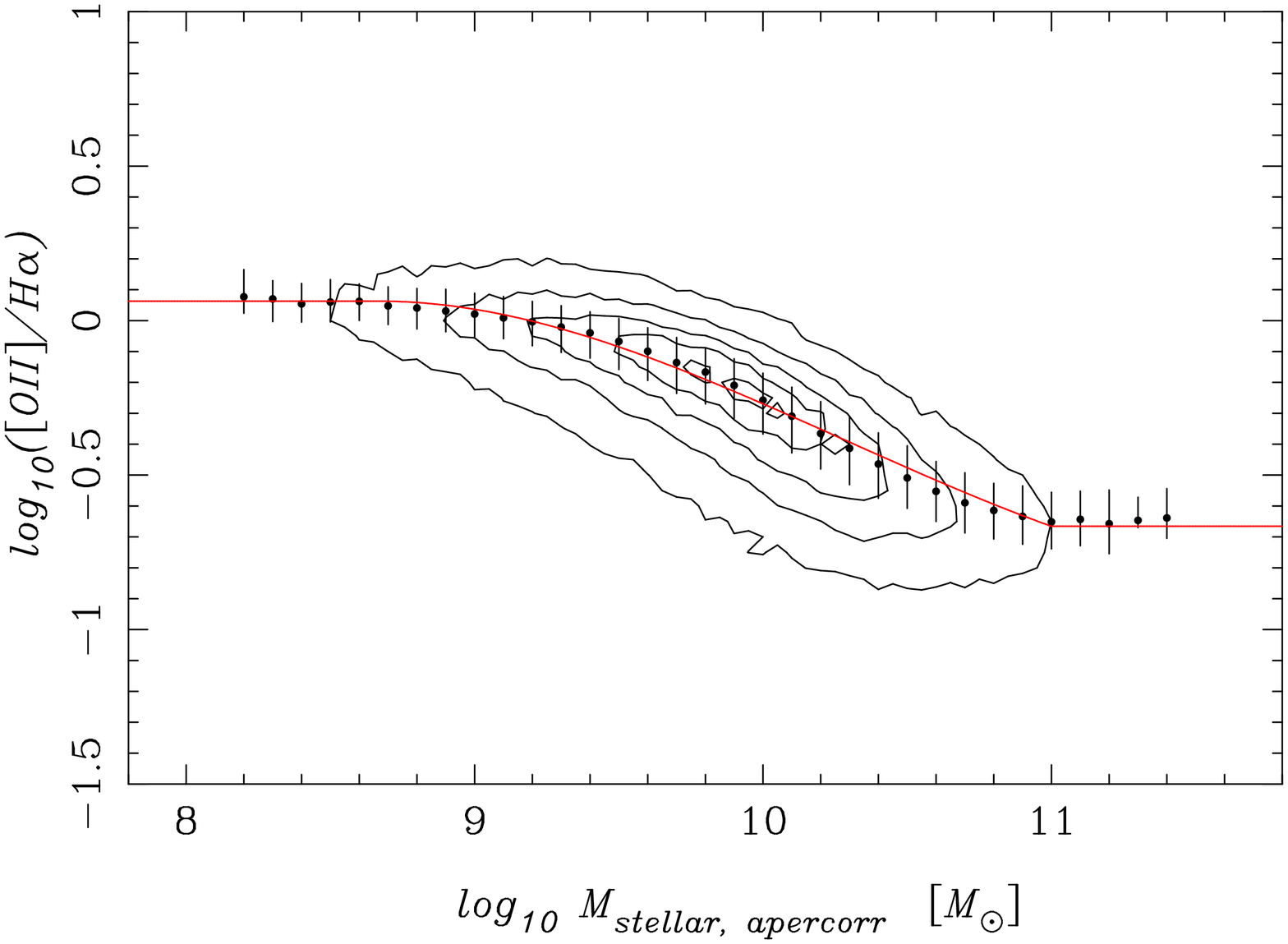}\hspace{0.5cm}
    \FigureFile(80mm,80mm){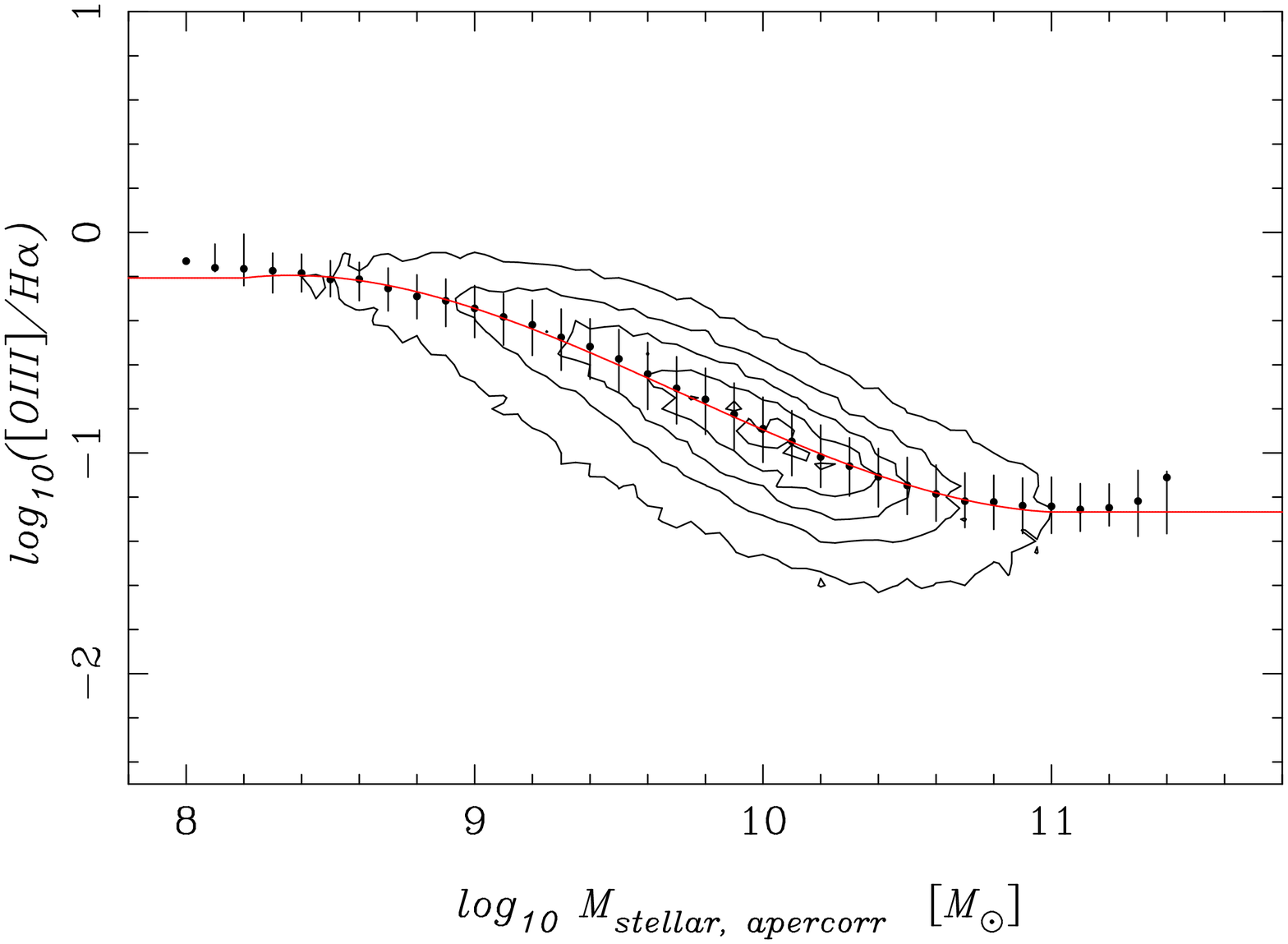}
  \end{center}
  \caption{
    {\bf Left:}
    Observed {\sc [oii]} to H$\alpha$ flux ratio plotted against stellar mass.
    The contours enclose 5\%, 25\%, 50\%, 75\% and 95\% of the galaxies.
    The points and error bars show the median and quartile of the distribution
    in each stellar mass bin.
    The curve shows the fitted relation.
    {\bf Right:} Same as the left panel, but {\sc [oiii]} to H$\alpha$ flux ratio
    is taken in the vertical axis.
  }
  \label{fig:oiii_ha_smass}
\end{figure*}

In addition to the Balmer lines, we make an extensive use of
{\sc [oii]} and  {\sc [oiii]} lines.
But, to derive {\sc [oii]} and {\sc [oiii]}, we need an empirical
calibration because these lines have strong dependence on metallicity,
but we cannot estimate metallicity of galaxies from the spectral fits
as discussed in the appendix.
We fit a relationship between
observed flux ratios of {\sc [oii]} or {\sc [oiii]} to H$\alpha$
and stellar mass as shown in Fig. \ref{fig:oiii_ha_smass}.
The idea is to use stellar mass as a proxy of metallicity given
the tight mass-metallicity relation \citep{tremonti04}.
Using these empirical calibrations, we can now predict {\sc [oii]} and
{\sc [oiii]} luminosity due to star formation (i.e., $L_{SF}$) as
we know the stellar mass and H$\alpha$ luminosity from the spectral fits.
We find that the predicted {\sc [oii]} and {\sc [oiii]} luminosities are slightly tilted with respect
to the observed luminosities (a log-linear slope of $\sim0.8$).
The tilt does not strongly affect our results in this paper and those in Paper-II
and for simplicity we assume that the predicted and observed luminosities of
all the emission lines are correlated with a log-linear slope of 1 with a small constant offset.

\subsection{Line of choice: H$\alpha$, {\sc [oii]}, or {\sc [oiii]}?}

\begin{figure*}
  \begin{center}
    \FigureFile(50mm,80mm){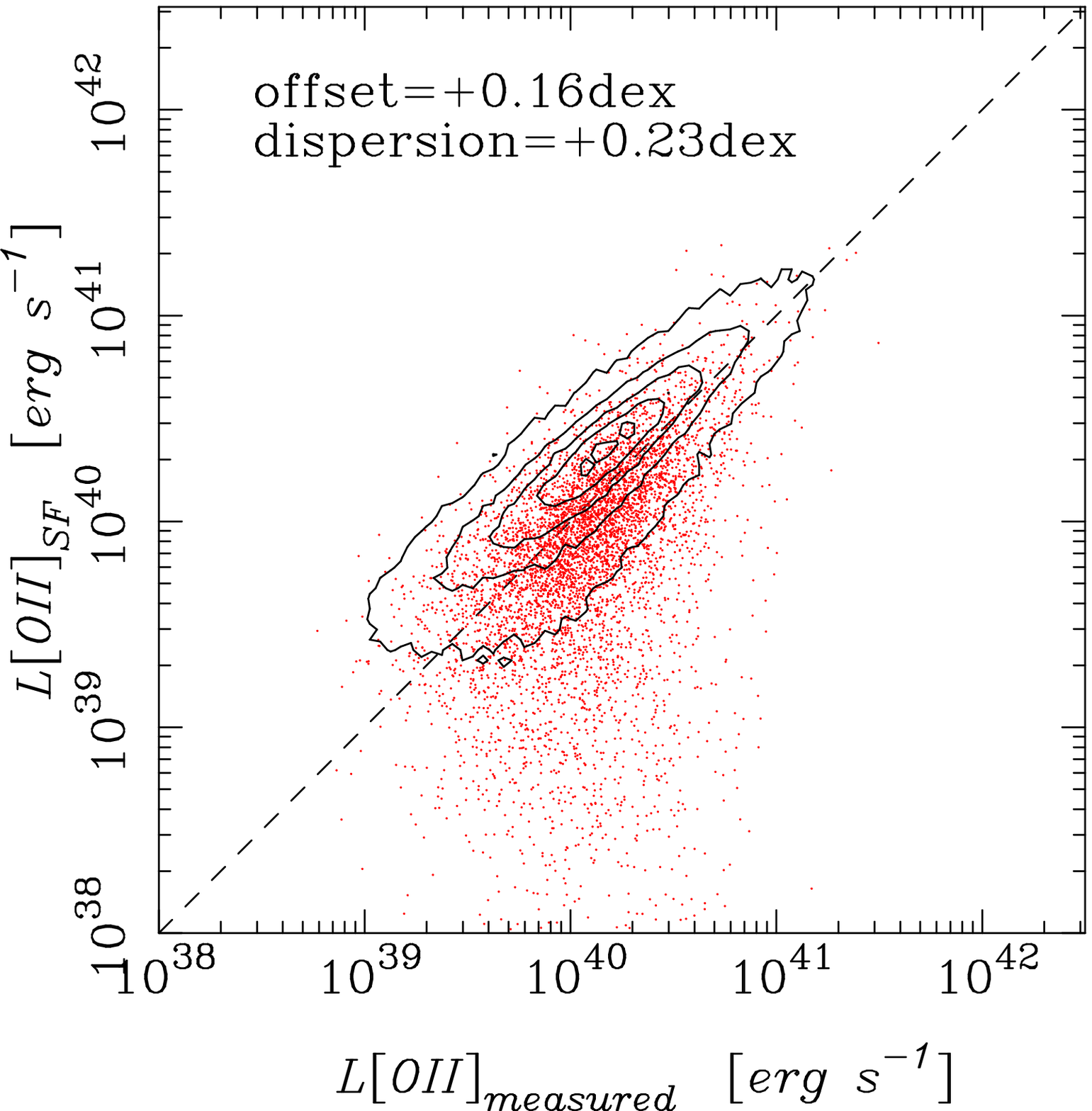}\hspace{0.5cm}
    \FigureFile(50mm,80mm){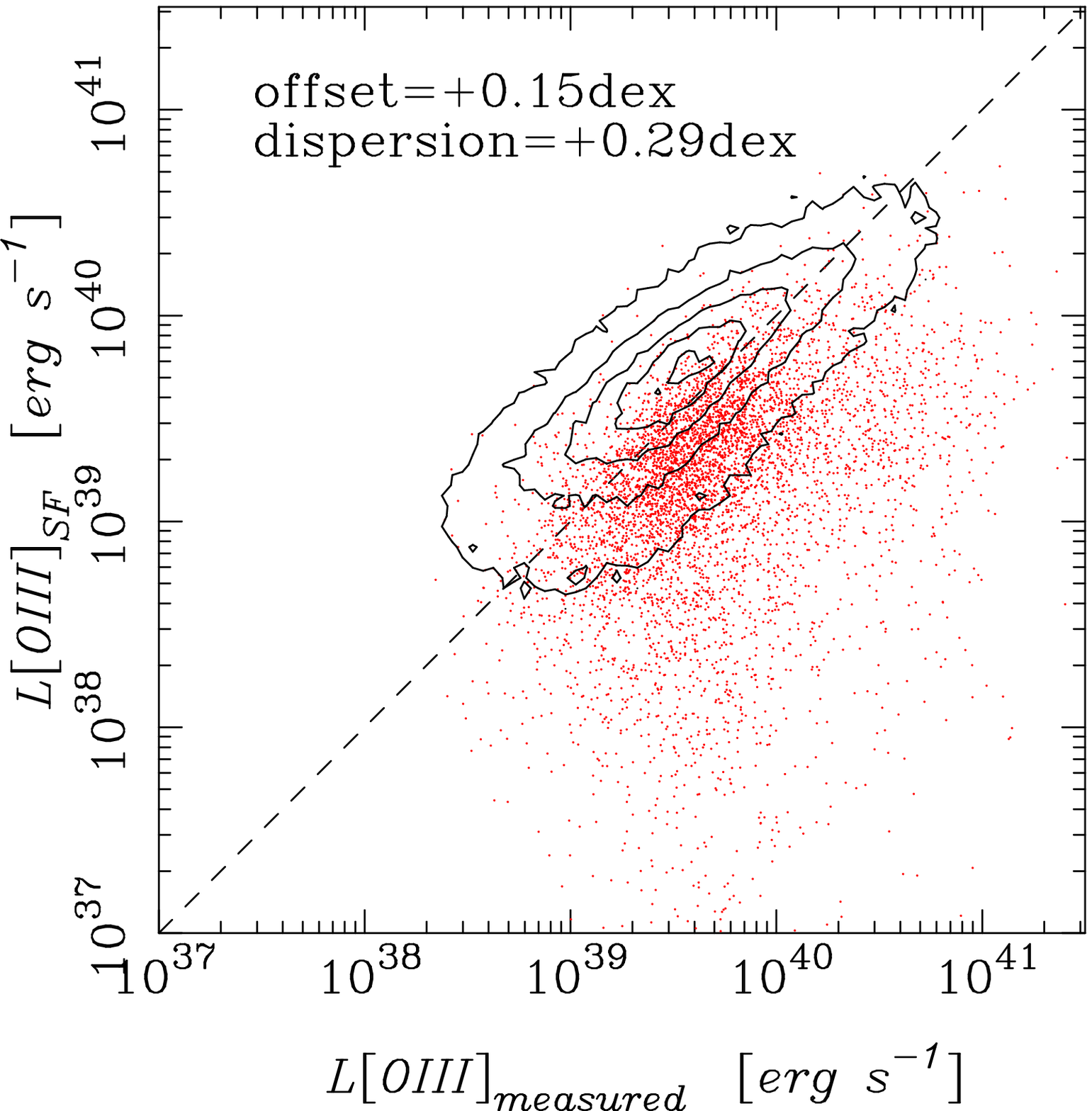}\hspace{0.5cm}
    \FigureFile(50mm,80mm){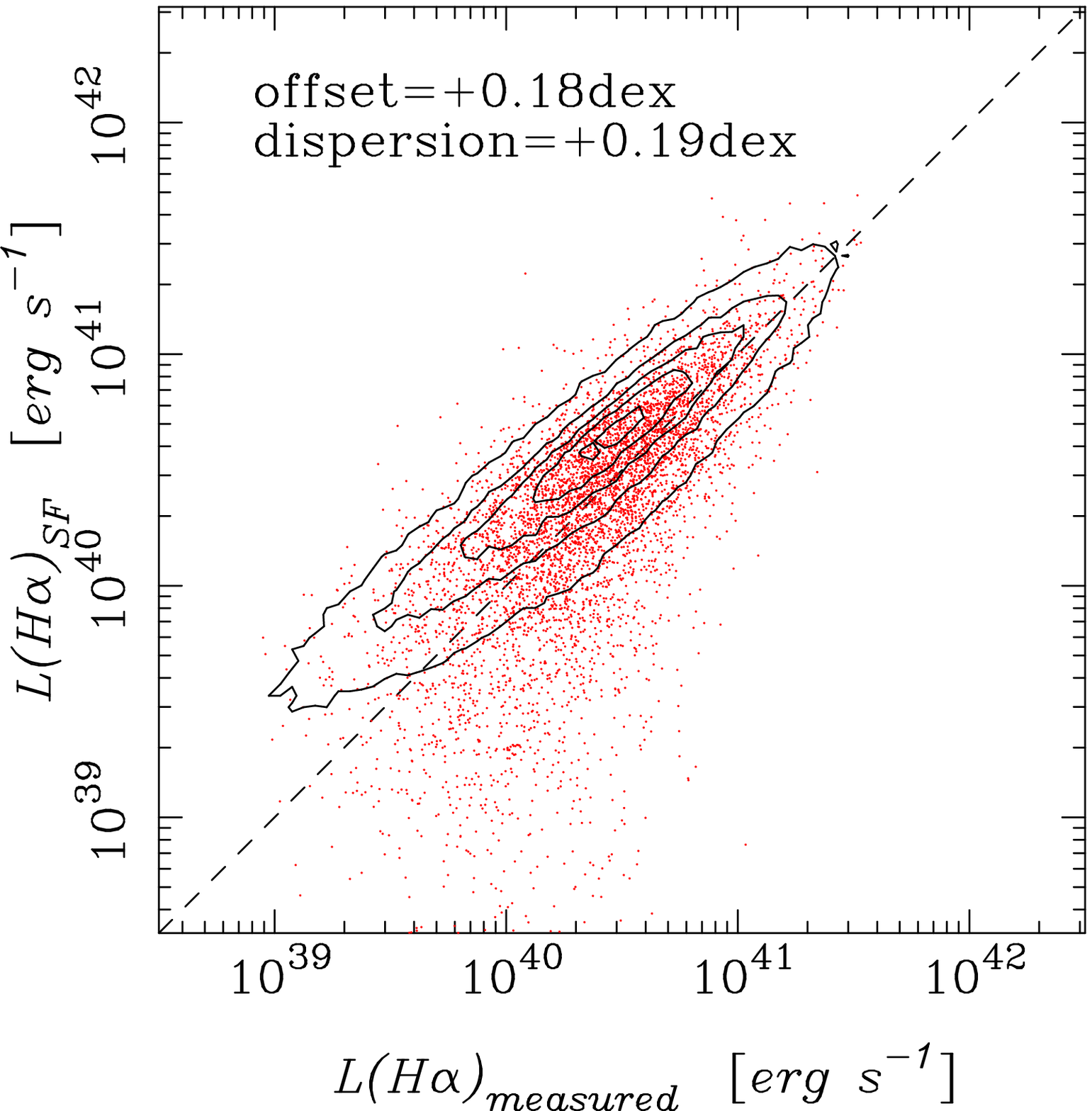}
  \end{center}
  \caption{
    Predicted line luminosities due to star formation plotted against measured line luminosities.
    The panels show, {\sc [oii]}, {\sc [oiii]}, and H$\alpha$ from
    left to right.  We show objects with significant line detections ($>3\sigma$).
    In each panel, the contours show star forming galaxies and
    the dots show AGNs selected from the BPT diagram (plotted every five objects).
    The dashed lines show $L_{SF}=L_{measured}$.
    An offset from the dashed line and dispersion in $\log (L_{SF}/L_{measured})$
    in each panel are for star forming galaxies.
  }
  \label{fig:comp_predictions}
\end{figure*}

Now we have developed a technique to predict $L_{SF}$ for 
the strong lines such as {\sc [oii], [oiii]}, and H$\alpha$.
Which line(s) should we use to identify AGNs?
We compare these three lines in Fig. \ref{fig:comp_predictions}.
We define AGNs using the BPT diagnostics with the \citet{kauffmann03}
threshold just as a guide line for now and show them with dots
in the figures.  Interestingly, the collisionally excited
Oxygen lines of AGN hosting galaxies show a clear offset with respect to
star forming galaxies, while H$\alpha$ line shows only a small offset.
This is what expected from the BPT diagram, which clearly shows
enhanced strengths of collisionally excited lines compared
to the Balmer lines in AGNs
(AGNs form a sequence towards the top-right corner of the BPT diagram,
not to the bottom-left corner).
In this work, we shall use {\sc [oii]} and/or {\sc [oiii]} for our purpose
of AGN identification.

As we will show in Paper-II, the ionization state of the narrow
line regions in AGNs spans a wide range.
Given this wide range of ionization, we deem that a sum of 
{\sc [oii]} and {\sc [oiii]} is likely a better indicator of
AGN activities than either one of them because these two lines balance
each other in gaseous nebulae and the sum of them is more robust
against variations in the ionization states.  We therefore use
the sum of the two Oxygen lines to identify and to characterize AGNs.

\begin{figure}
  \begin{center}
    \FigureFile(80mm,80mm){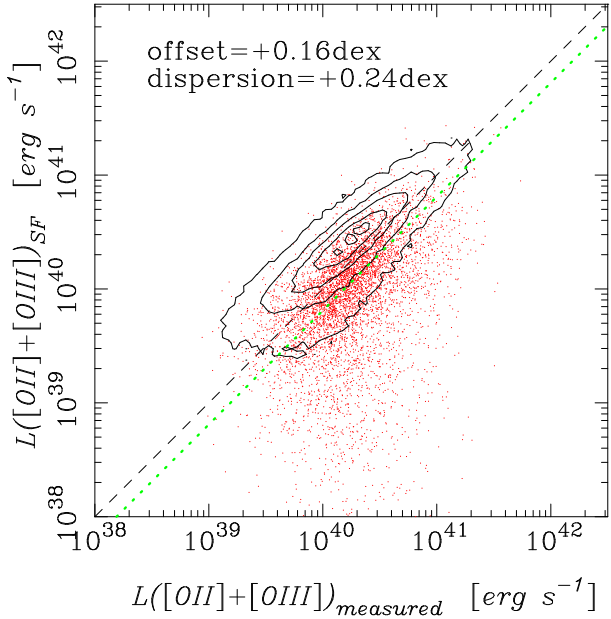}
  \end{center}
  \caption{
    Predicted {\sc [OII]+[OIII]} luminosities due to star formation
    plotted against measured luminosities.
    The contours show star forming galaxies and
    the dots show AGNs selected from the BPT diagram.
    Only every 5 AGNs are plotted for clarity.
    The dotted line shows the threshold of Oxygen-excess and
    galaxies below it are defined as Oxygen-excess galaxies.
  }
  \label{fig:oiioiii_comp}
\end{figure}

Using the technique developed above, we can predict 
the {\sc [oii]+[oiii]} luminosities fairly well
as shown in Fig. \ref{fig:oiioiii_comp}.
We define {\it Oxygen-excess galaxies} as those with

\begin{eqnarray}
\label{eq:agn_def}
\log_{10} \frac{L_{[OII]+[OIII], measured}}{L_{[OII]+[OIII],SF}} - \log_{10} L_{[OII]+[OIII],offset} \nonumber \\
  > 1.5\times\log_{10} \sigma(L_{[OII]+[OIII], SF}),  \hspace{0.5cm}
\end{eqnarray}

\noindent
where $L_{[OII]+[OIII], offset}$ is a systematic offset between the predicted
luminosity and measured luminosity for star forming galaxies ($+0.16$ dex as
shown in  Fig. \ref{fig:oiioiii_comp}) and $\sigma(L_{[OII]+[OIII], SF})$
is an accuracy of our luminosity predictions ($0.24$ dex).  Our selection 
criterion is that if an object has an {\sc [oii]+[oiii]} luminosity
that exceeds the expected luminosity due to star formation by $>1.5\sigma$,
it is an Oxygen-excess object.
This sigma cut is a trade off between completeness and contamination (e.g.,
if we reduce the threshold to $1\sigma$, we have a better sampling of BPT AGNs
at the cost of increased contamination of star forming galaxies).
Our choice of $1.5\sigma$ is simply a compromise between them, but we have confirmed
that our results do not significantly change if we change it to $1\sigma$ or $2\sigma$.
The adopted threshold is shown as the dotted line in Fig. \ref{fig:oiioiii_comp}.
We further require a significant detection of {\sc [oii]+[oiii]} at $>3\sigma$
to ensure that we do not suffer from noises.
Since the idea is to identify Oxygen emission line excess,
we dub our method ``Oxygen-excess method''.

\begin{figure*}
  \begin{center}
    \FigureFile(80mm,80mm){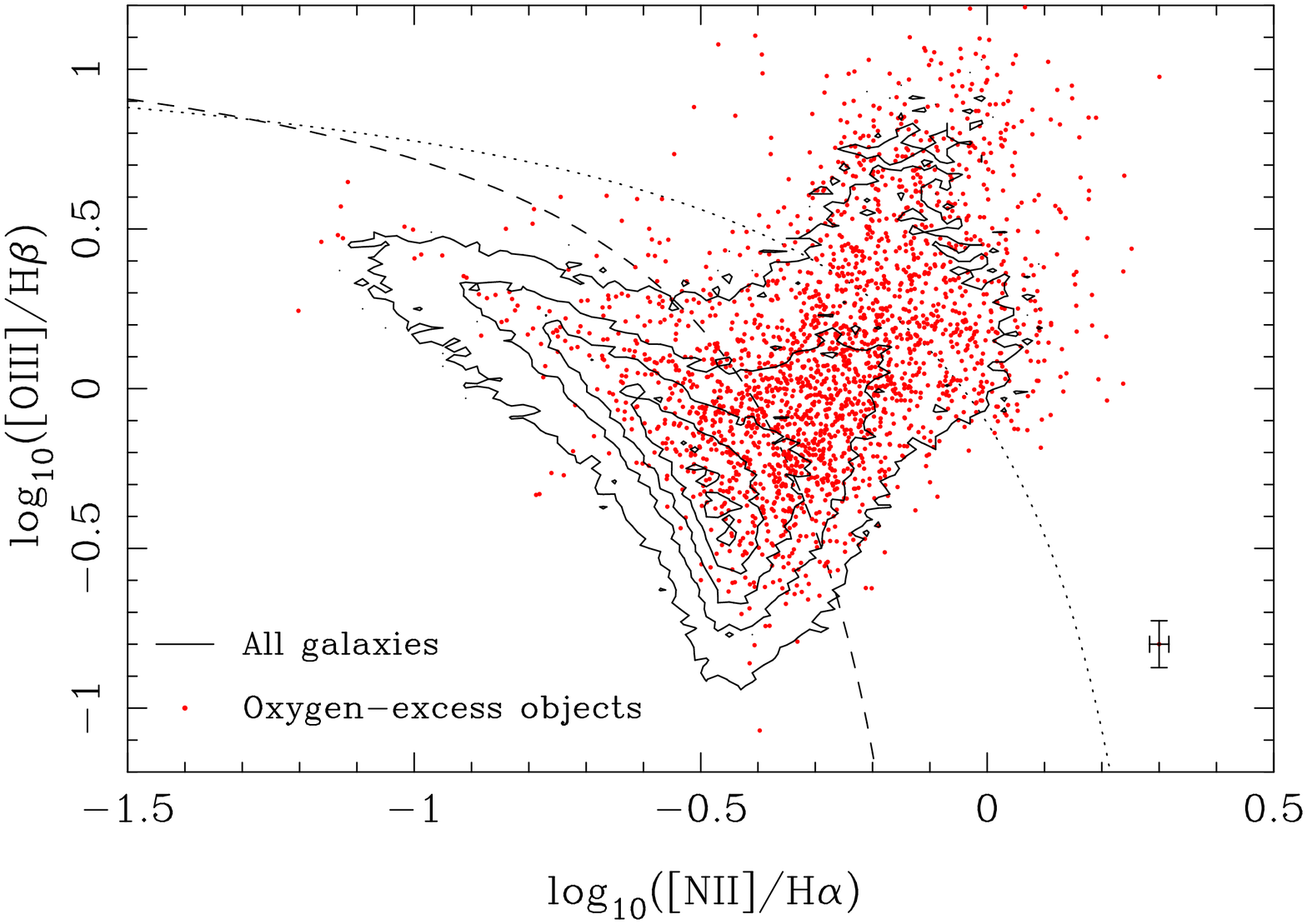}\hspace{0.5cm}
    \FigureFile(80mm,80mm){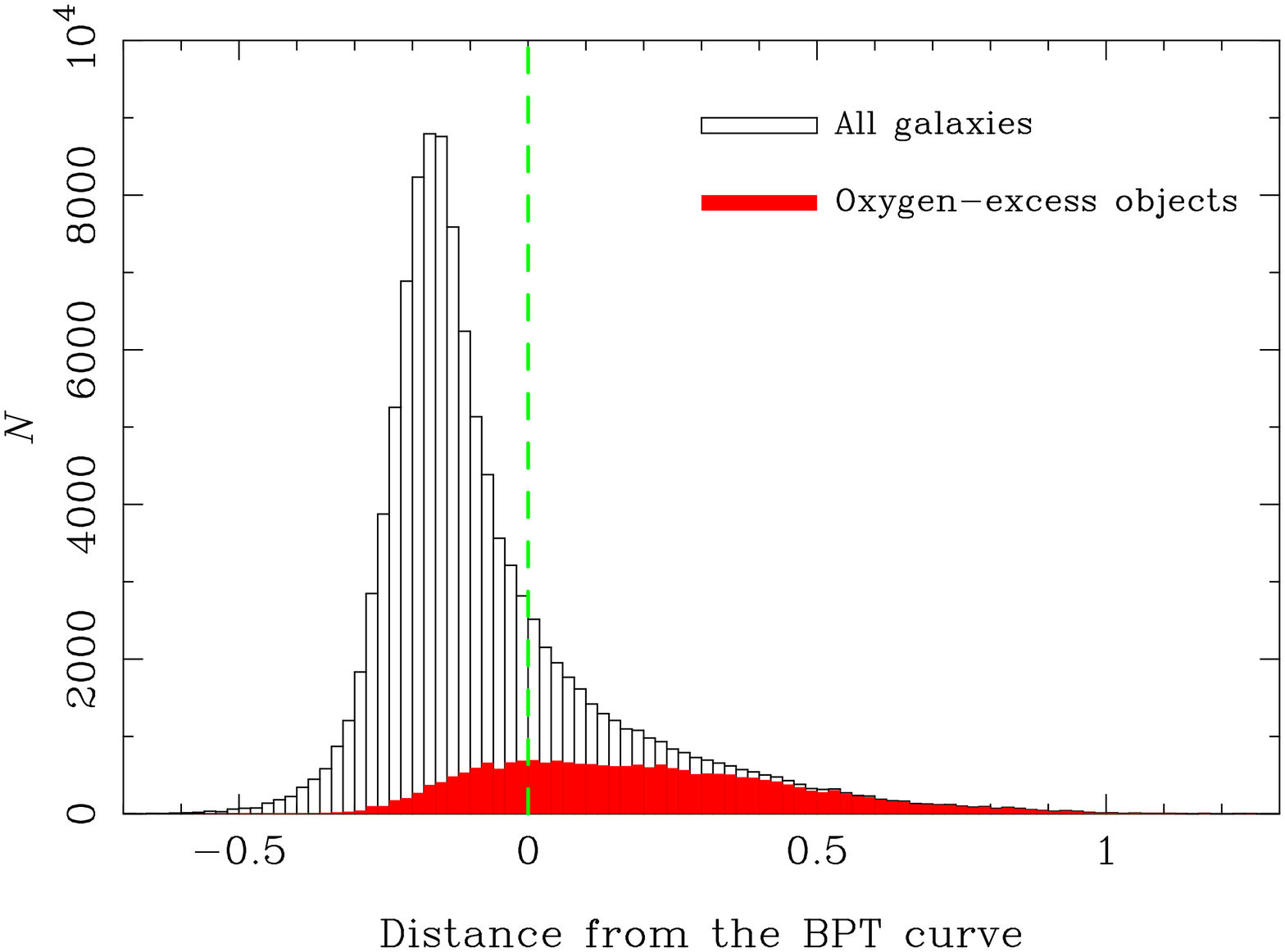}
  \end{center}
  \caption{
    {\bf Left:}
    BPT diagram.
    The contours encircle 5\%, 25\%, 50\%, 75\% and 95\% of all the galaxies
    used here (i.e., $0.02<z<0.1$ and all the four lines detected at $>3\sigma$).
    The dots show Oxygen-excess objects (plotted every five objects). 
    The dashed curves is from \citet{kauffmann03}.
    Just for a reference, we also show the \citet{kewley01} threshold
    as the dotted curve.
    The cross on the bottom-right shows the median error in the line ratios.
    {\bf Right:}
    The distribution of 'distance' of galaxies from the \citet{kauffmann03} curve in the left panel.
    The negative distance means that galaxies are in the star forming region, while
    the positive distance means that galaxies are in the AGN region.
    The vertical dashed line is the \citet{kauffmann03} threshold.
    The open and filled histograms show all the galaxies and 
    Oxygen-excess galaxies.
  }
  \label{fig:bpt_diagram}
\end{figure*}

As a quick check of the Oxygen-excess method developed here,
we plot in Fig. \ref{fig:bpt_diagram}
the distribution of the Oxygen-excess galaxies on the BPT diagram.
43\% of galaxies in our sample show strong enough emission lines
to apply the BPT diagnostics and are plotted as the contours in the figure.
Among these strong emission line objects,
we find that our SF/AGN classifications agree with BPT for 85\% of objects.
The Oxygen-excess objects are mostly (75\%) in the AGN region of the BPT diagram.
The rest of them are in the SF region, but
we note that the SF/AGN threshold of \citet{kauffmann03} is arbitrary defined.
If we use the criterion proposed by \citet{stasinska06},
the fraction increases to 87\%.

The right plot of Fig. \ref{fig:bpt_diagram} shows 
the distribution of projected distances of the galaxies to the threshold curve.
Galaxies distribute contiguously around the (arbitrary set) threshold.
This continuous sequence from star forming galaxies
to AGNs probably represents a wide range of AGN activities with
respect to underlying star formation.
The BPT diagnostics misses weak AGNs in actively star forming galaxies
(so does our method but with a improved sensitivity to weak AGNs;
see the next section and Paper-II)
and it would not be surprising at all if some of the galaxies
in the star forming region of the diagram actually host AGNs.
In fact, the Oxygen-excess objects in the star forming region of the diagram
are skewed towards $distance\sim0$, while normal star forming galaxies
form a peak around $distance\sim-0.15$.
If these Oxygen-excess objects were pure contamination, we would
have seen a peak at 
$distance\sim-0.15$ with an extended tail to $distance\sim-0.5$.
The skewed distribution of the Oxygen-excess objects suggests that
they are not pure contamination of star forming galaxies.
We will make further attempts to characterize our method
in the next section.

In addition to the most commonly used BPT diagnostics, we also present
{\sc [oiii]}/H$\beta$ vs. {\sc [oi]}/H$\alpha$ and 
{\sc [oiii]}/H$\beta$ vs. {\sc [sii]}/H$\alpha$ diagrams \citep{veilleux87}
in the Appendix for completeness.
As pointed out by earlier papers (e.g., \cite{kewley06,stasinska06}),
the separation between star formation and AGN sequences is less clear
if we use {\sc [oi]} or {\sc [sii]} in place of {\sc [nii]}.
{\sc [nii]} seems to work better because the {\sc [nii]}/H$\alpha$ ratio
saturates at high metallicities \citep{kewley02} due likely to its
secondary nature \citep{kewley02,stasinska06}.
But, the {\sc [oi]} and {\sc [sii]} lines are interesting in their own right.
For instance, the {\sc [oi]} line comes from partially ionized nebulae and
hence it is sensitive to the hardness of ionizing spectrum
and is a good probe of AGNs.
Readers are refereed to Appendix for further discussions.

Finally, we shall emphasize that the Oxygen-excess method can be applied
at much higher redshifts compared to BPT.
Although {\sc [oii]+[oiii]} is the most effective set of lines,
one can in principle use any line to identify an emission line excess.
A practical application of the method would be to use {\sc [oii]} only.
One has to take a risk of missing high ionization AGNs that exhibit weak {\sc [oii]},
but the gain is that one can go up to $z\sim1.7$ with red-sensitive optical spectrograph
to study the AGN evolution over a wide redshift range.
Our method will be an ideal method to identify AGNs and study their host galaxy
properties in on-going/near-future massive spectroscopic surveys such as
SDSS-III Baryon Oscillation Spectroscopic Survey.

\section{Comparisons between Oxygen-excess, BPT, X-ray and radio sources}

Following the development of the Oxygen-excess method,
we make extensive tests of the method
by comparing with other AGN detection methods in this section.
  First, we further compare with
the BPT diagnostics.    A significant fraction of galaxies 
in our sample ($43$\%) show too weak
emission lines to apply the BPT diagnostics, but they have strong enough Oxygen lines
to apply the Oxygen-excess method.  We test how well our method works
in such weak emission line galaxies by stacking spectra.
The stacked spectra are also useful to characterize average properties
of various classes of objects.
We then compare with X-ray sources identified in
archival Chandra observations and also with radio sources from FIRST.

Before we present our results, it is important to emphasize that
none of the Oxygen-excess, BPT, X-ray, and radio methods is a perfect
method to identify AGNs.
Each method has pros and cons and they all suffer from incompleteness
and contamination.
The most relevant numbers to quote in this section would be fractions of
missing AGNs and contaminating non-AGNs in the Oxygen-excess objects.
But, we are unable to provide these numbers because no AGN identification
method gives a complete sampling of AGNs.  Nonetheless, we make an attempt
to quantify whether a majority of the Oxygen-excess are real AGNs or not.
Note that
we cannot reach any clear conclusion in intermediate types of objects that
are fundamentally difficult to classify (see \cite{ho08} for a review of the subject).
Also, we will discuss objects that are photo-ionized by non-AGN sources in section 3.4.

\subsection{Stacked objects on the BPT diagram}

\begin{table*}[t]
  \begin{center}
    \caption{
      Galaxy population classes.
    }
    \label{tab:agn_class}
    \begin{tabular}{cccc} 
                          &  BPT-AGN  &  BPT-SF       & 4 lines unavailable\\\hline
      Oxygen excess       &  $O+B+$   &  $O+B-$       & $O+Bn$    \\
      No Oxygen excess    &  $O-B+$   &  $O-B-$       & $O-Bn$    \\
      Oxygen unavailable  &  ---      &  ---          & $OnBn$    \\
    \end{tabular}
  \end{center}
\end{table*}

We define classes of galaxy populations in Table \ref{tab:agn_class}
to compare the Oxygen-excess objects with the BPT objects.
The notation is defined as 
'O' and 'B' stand for Oxygen-excess and BPT.
'+' and '-' mean AGN and SF.  For example, O+B- are the objects
that are identified as AGN by the Oxygen-excess method, but are classified
as SF by BPT.
We use an 'n' for objects we cannot apply the Oxygen-excess or BPT
method due to weak emissions.
Note that we use the threshold by \citet{kauffmann03}
to define AGN and SF on the BPT diagram.

In our spectral fitting described in the last section, we subtract
the continuum using the best-fitting model template and further
by applying the median filter to remove the residuals.
We stack these continuum subtracted spectra in each class
using the inverse-variance weights to make the average
emission line spectra.
We also perform the median stacking in addition to the
inverse-variance stacking.
The emission line strengths in the stacked spectra are somewhat
different between these two stacking techniques, but the line ratios,
which we will soon discuss, are not very different.
Note that we combine the spectra in apparent flux density, not in
distance corrected luminosity density (i.e., only the wavelengths are corrected
to rest-frame) because the AGN/SF classification is
limited by the observational flux limit (this is especially relevant to
O+Bn, O-Bn, and OnBn classes) and we would like to show typical observed flux densities of
objects in these classes and compare them with stronger emission line objects.
We have confirmed that the line ratios we discuss below are
essentially the same regardless of whether we correct for
the distance or not.

\begin{figure*}
  \begin{center}
    \FigureFile(170mm,80mm){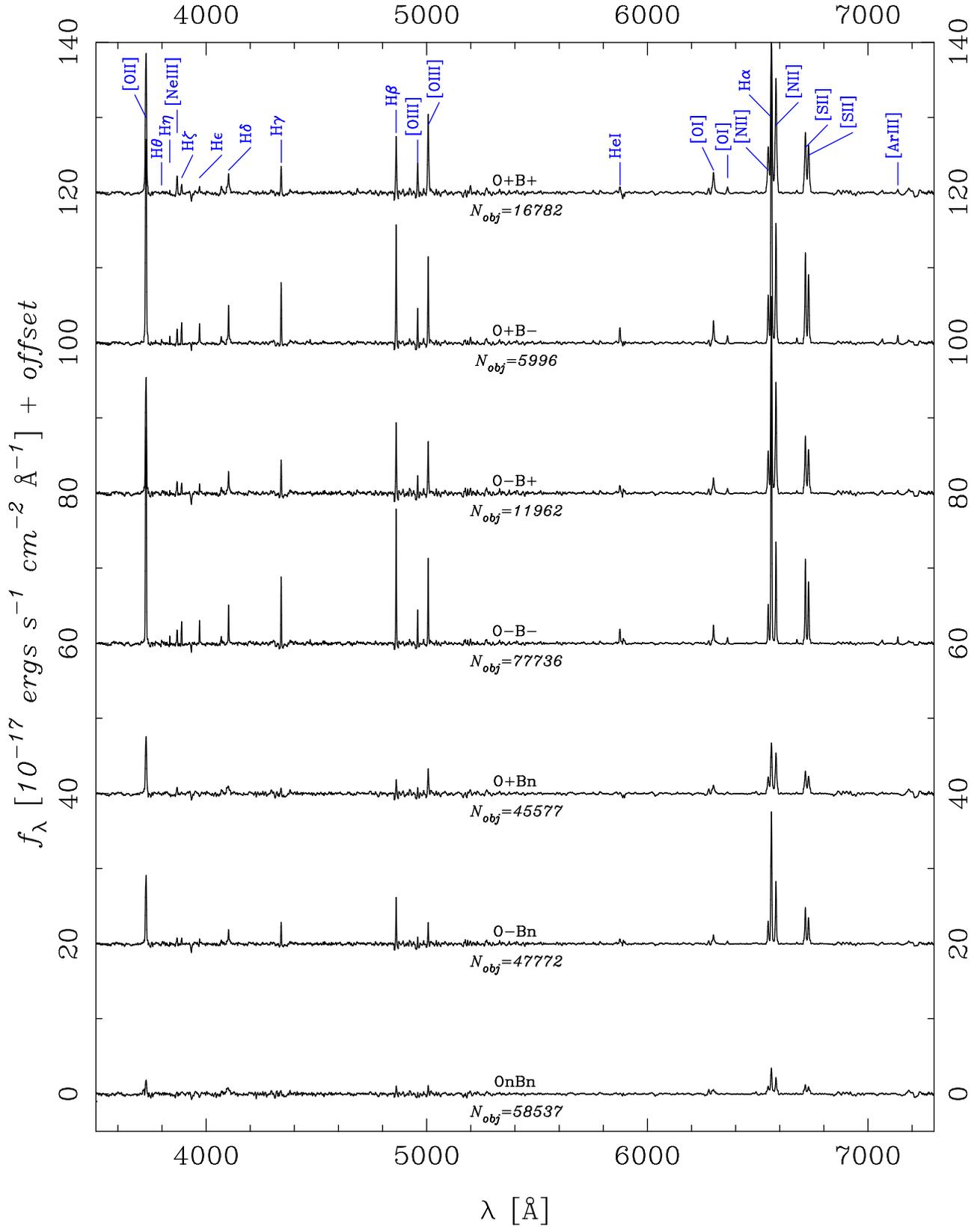}\\
  \end{center}
  \caption{
    Stacked spectra.  From top to bottom, the spectra are for
    O+B+, O+B-, O-B+, O-B-, O+Bn, O-Bn, and OnBn, respectively.
    Most prominent emission lines are labeled.
    The numbers of objects used for the stacking are shown as well.
  }
  \label{fig:stacked_spectra}
\end{figure*}

\begin{figure*}
  \begin{center}
    \FigureFile(120mm,80mm){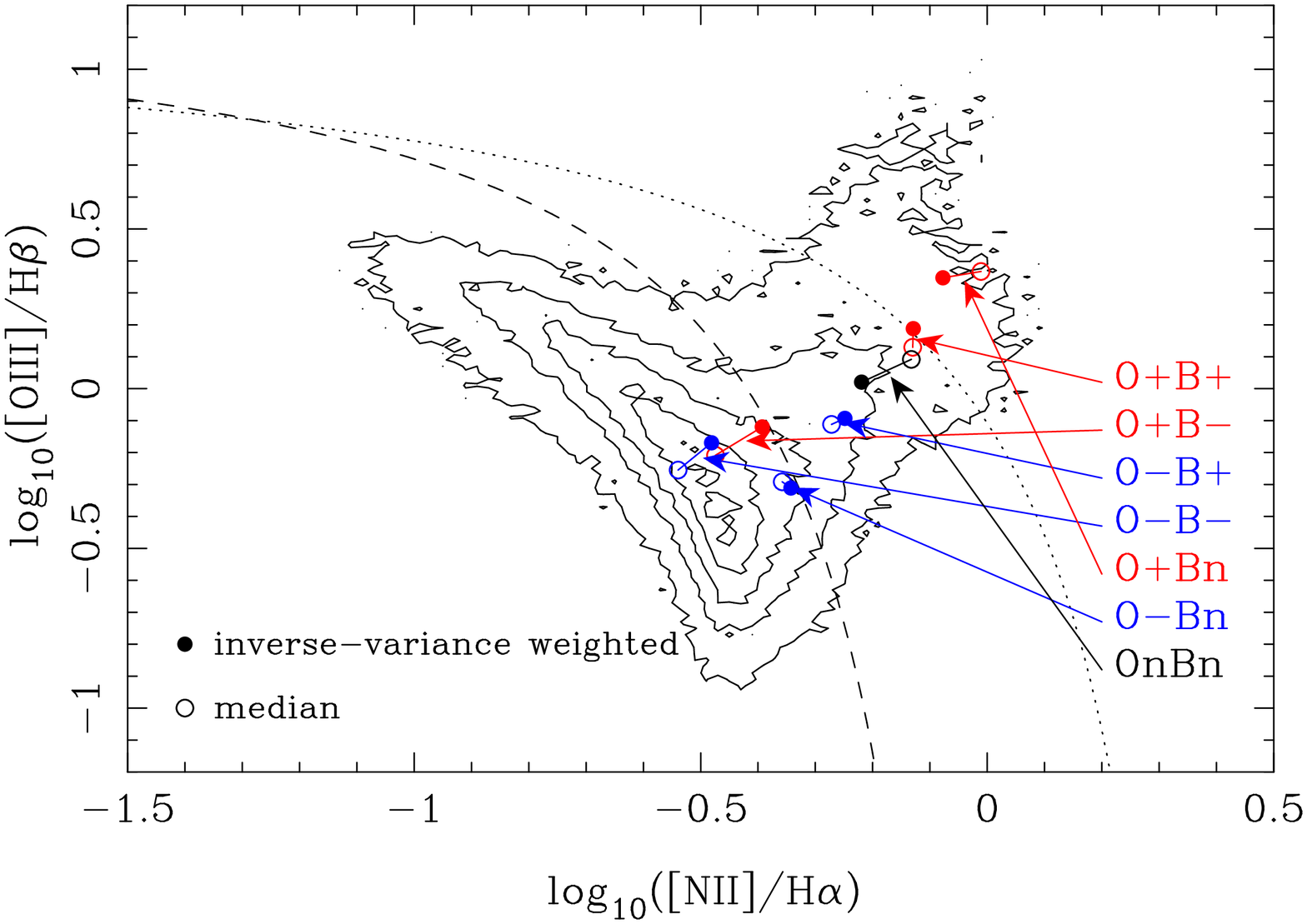}\hspace{0.5cm}
  \end{center}
  \caption{
    BPT diagram.  The contours show galaxies with strong emission lines as in Fig. \ref{fig:bpt_diagram}.
    The locations of the stacked objects are
    indicated by the circles and arrows. The filled and open
    circles are measured from the inverse-variance weighted stacking
    and from the median stacking, respectively.
  }
  \label{fig:bpt_diagram_stacked}
\end{figure*}

Fig. \ref{fig:stacked_spectra} presents the stacked spectra
of all classes.
Note the very high quality of the stacked spectra.
A typical emission line fluxes in the OnBn class is comparable
to the typical noise level of the SDSS spectra.
The stacked spectra are of sufficient signal-to-noise
to measure emission line fluxes for all class
of objects.
We can now measure the line ratios and study their average
properties.
O+Bn and O-Bn will be particularly interesting because
we cannot apply the BPT diagnostics to these galaxies individually.
We show locations of the stacked objects on the BPT diagram
in Fig. \ref{fig:bpt_diagram_stacked} and discuss each class
of objects below.
The numbers in the parenthesis are the fractions
of objects in that class to the entire sample.

\begin{itemize}
\item {\bf O+B+ (6.3\%):}
We observe strong emission lines in the stacked spectra.
The line flux ratios on the BPT diagram indicate that the galaxies host AGNs.

\item {\bf O+B- (2.3\%):}
This is an interesting class of objects.  The galaxies are
classified as star forming galaxies on the BPT diagram, but
we observe an Oxygen flux excess.
The stacked spectrum
ranges from the middle of the star forming sequence to the threshold
line of \citet{kauffmann03}.
The observed offset to the threshold compared to B-O-, which are
nearly pure star forming galaxies (see below), suggests that
these galaxies may harbor weak AGNs with underlying active star formation.
But, the BPT diagram does not give us an estimate of
AGNs and contaminating star forming galaxies.
This is one of the difficult classes to characterize.

\item {\bf O-B+ (4.5\%):}
The galaxies are defined as AGNs from the BPT diagram, but we do
not observe a significant Oxygen flux excess.
The stacked spectra and the location on the BPT diagram
show that these galaxies likely host AGNs with active underlying star formation.
The X-ray analysis below also suggests that they are likely AGNs.
The majority of the galaxies in this class are AGNs and our method misses them.
It could be that we miss them due to statistical fluctuations of
our flux predictions (the scatter is a factor of 1.7; Fig. \ref{fig:oiioiii_comp}).
But, it could also be that AGN continuum gives a non-negligible
contribution to the overall continuum spectra.
AGN continuum is likely a power-law form in the optical
wavelengths, making the spectra bluer.
Although such AGN continuum is typically fairly weak in weak AGNs \citep{schmitt99},
we may over-estimate SFRs if AGN continuum is happen to be strong.
As a result, we may miss AGNs.
As mentioned earlier, we cannot measure the amount of
such continuum contribution with our spectral fits well.
GALEX photometry may give us an insight into AGN continuum,
but our poor estimates of dust extinction would not allow
us to study near-far UV luminosities because of the
strong sensitivity to dust.  A more sophisticated spectral
fitting would be needed.
We characterize the host galaxy properties of O-B+ in the Appendix
and show that this missing AGN population does not affect
our conclusions in Paper-II.

\item {\bf O-B- (29.4\%):}
These objects are in the star forming region on the
BPT diagram and we do not observe an Oxygen flux excess.
The stacked spectrum is indeed in the middle of
the star forming region of the BPT diagram.
Therefore, these objects are likely star forming galaxies.
\end{itemize}

We can apply the BPT diagnostics to all the objects discussed so far\footnote{
If we use the \citet{kewley01} threshold to define AGN/SF,
the fractions in the above classes are 2.5\%, 5.8\%, 0.3\%, 32.5\% for
O+B+, O+B-, O-B+, and O-B-, respectively.
}.
We shall emphasize that the SF/AGN classification by the Oxygen-excess
method is consistent with BPT for 85\% of the objects (O+B+ and O-B-).
Intermediate cases (O+B- and O-B+) are somewhat challenging to fully characterize,
but they make up only 15\% of the strong emission line objects.
{\bf
}
We also emphasize that we could apply the BPT diagnostics only to
43\% of objects in our sample.
Although we define our sample at $0.02<z<0.10$, which is below the median redshift of the Main galaxy
sample ($z\sim0.1$), more than a half of the objects remain unclassified.
This is where our method has a great advantage --- our method can be applied to
nearly twice as many objects as BPT.  We discuss these weak emission
line objects below.

\begin{itemize}
\item {\bf O+Bn (17.2\%):} 
The galaxies in this class do not show strong enough
lines to apply the BPT diagnostics, but their {\sc [oii]+[oiii]}
lines are strong enough to apply our method.
A large fraction of galaxies in the O+Bn and O-Bn classes demonstrates
the sensitivity of our method to low-luminosity objects.
The emission lines in the stacked spectrum are weak as expected.
H$\beta$ is often the weakest line among the four lines used
in the BPT, and that limits the sensitivity of BPT to low-luminosity AGNs.
The line ratios from the stacked spectra clearly show that these objects
actually host AGNs.  This is a strong proof that the Oxygen-excess method
works well in identifying such low-luminosity AGNs.
The X-ray and radio analyses presented below also shows that
most of the objects are likely AGNs.
We shall note that there has been a considerable debate as to whether
low-luminosity, low-ionization objects (LINERs; \cite{heckman80}) are powered by AGNs.
There are other energy sources proposed in the literature
that can produce weak LINER-like emission.  We discuss those fake AGNs
in Section 3.4.

\item {\bf O-Bn (18.1\%):}
The stacked spectrum shows weak emission lines.
H$\beta$ is stronger than {\sc [oiii]}
and this class of objects unlikely host strong AGNs.
The stacked object lies on the border of the threshold line
in the BPT diagram.  This suggests that, while the majority of the objects
do not host AGNs, we may have small contamination of AGNs
(see the radio analysis below).
O+Bn and O-Bn are the classes for which we can apply the Oxygen-excess method only.
The positions of these objects on the BPT diagram clearly demonstrate that
our classification for such weak emission line objects is fairly reasonable.

\item {\bf OnBn (22.1\%):}
A large fraction of all the objects do not show even weak emission lines
and fall in this category.
If we stack their spectra, very weak emission lines emerge.
Most of the lines are comparable strength to
typical noise level in the SDSS spectra.  The line ratios in Fig.
\ref{fig:bpt_diagram_stacked} indicate
that LINERs reside in such apparently quiescent galaxies.
We also observe
that a fraction of them is detected in radio despite their weak emission (see below).
\end{itemize}

\begin{figure}
  \begin{center}
    \FigureFile(80mm,80mm){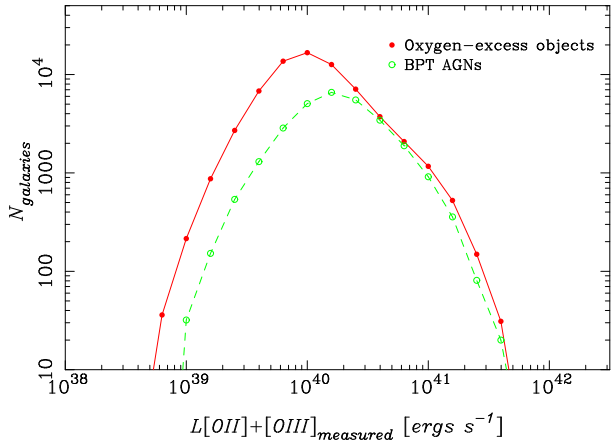}\hspace{0.5cm}
  \end{center}
  \caption{
    $L_{[OII]+[OIII]}$ distributions of the Oxygen-excess objects (filled circle)
    and BPT AGNs (open circles).
  }
  \label{fig:agn_comp5}
\end{figure}

To sum up, for bright objects for which we can apply the BPT diagnostics,
the Oxygen-excess method gives the consistent SF/AGN classifications
for 85\% of the objects.
While the BPT diagnostics is applicable to $43\%$ of all the objects,
the Oxygen-excess method is applicable to $78\%$.
This results in a higher fraction of the Oxygen-excess objects (26\%)
than BPT AGNs (11\%).
Fig. \ref{fig:agn_comp5} illustrates our sensitivity to low-luminosity objects.
The Oxygen-excess method can identify a significant number of low-luminosity
objects that are missed by BPT.
The stacked spectra show the average properties of objects in each class.  
Except for a small fraction of the discrepant cases (O+B- and O-B+,
$7\%$ of total), the stacked spectra on the BPT diagram show
that the AGN/SF classifications are made well.
In particular, low-luminosity objects for which the BPT is not
applicable are classified fairly well.
This is an encouraging result and motivates us to perform
a further test of the method using X-ray and radio data.

\subsection{X-ray sources}

\begin{figure}
  \begin{center}
    \FigureFile(80mm,80mm){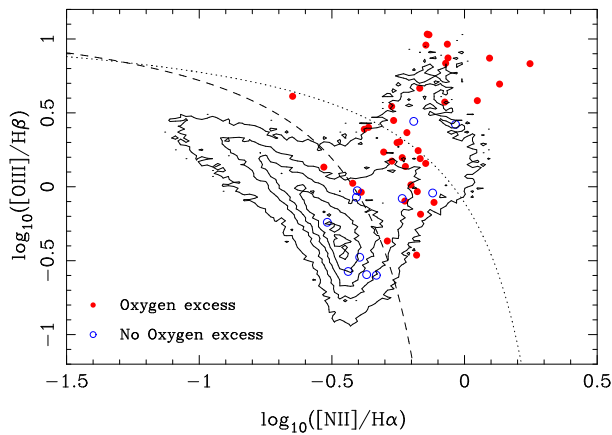}\hspace{0.5cm}
  \end{center}
  \caption{
    Distribution of X-ray sources on the BPT diagram.
    The curves and the contours are as in Fig. \ref{fig:bpt_diagram}.
    The filled and open points show X-ray sources with and without
    Oxygen flux excess, respectively.
  }
  \label{fig:bpt_xray}
\end{figure}

We base our X-ray analysis on the CSC-SDSS cross-match catalog
release 1.1 from the Chandra website\footnote{http://cxc.harvard.edu/csc/}
\citep{evans10}.
The catalog has Chandra sources from archival data matched to SDSS objects.
Due to the nature of its serendipity, the data depth varies across
the sky and the data are not defined in any systematic way.
However, the serendipity in turn
allows us to perform a statistical test of our method because any specific
types of Oxygen-excess objects do not prefer any particular patch of Chandra
observations.

We use SDSS objects that have point-like Chandra counterparts within 2 arcsec
from the center to ensure that we are looking at sources at the galaxy centers,
while accommodating the PSF degradation at large angle from the focal axis.
This results in 56 X-ray sources, which are
the subject of the analyses in this subsection.
We have also cross-matched the archival XMM-Newton sources \citep{watson09}.
We find that the conclusions in this section remain the same if we use
the XMM sources, but
there seems a slightly increased amount of non-AGNs sources possibly
due to the poorer angular resolution of XMM compared to Chandra\footnote{
We have cross-matched 108 objects with the archival XMM sources within
1$\sigma$ positional error (but we impose a maximum separation of 3 arcsec).
We find that a larger fraction of objects have soft X-ray emission,
whose origin could both be AGN and non-AGN sources, compared to the Chandra sources.
Although we restrict the sources to be consistent with PSF sizes,
the galaxies optically extend on a comparable angular size to the 
resolution of XMM.  We suspect that the XMM luminosities suffer from
increased contamination from non-AGN sources such as 
X-ray binaries than Chandra.
}.
We thus use the Chandra data for the analysis here.

We do not correct for absorption due to the Galactic neutral hydrogen
because essentially all the objects have a low hydrogen column density due
to the Galaxy of a few times $10^{20}\rm cm^{-2}$ \citep{dickey90}.
Based on \citet{morrison83}, we estimate that the absorption in
the hard band (2-7keV) is negligible.  The absorption is still small
in the soft band (0.5-1.2keV) with an optical depth of a few times $10^{-1}$.
The correction for intrinsic absorption requires sufficient X-ray photons
and a detailed spectral analysis, which is beyond the scope of this work.
However, an intrinsic absorption is likely below $10^{22}\rm\ cm^{-2}$
at the X-ray luminosity range we explore here \citep{mainieri07}.
At this column density, a hard X-ray luminosity is largely unaffected by absorption.
Our hard X-ray luminosity should therefore be reasonable estimates.

First, we put the X-ray objects in the BPT diagram in Fig. \ref{fig:bpt_xray}.
Note that we cannot show X-ray detected O+Bn, O-Bn, and OnBn objects in the diagram
due to their weak emission.
Most of the X-ray sources are in the AGN region of the diagram.
There are several sources in the star forming region, 
but most of them do not show any Oxygen excess.

\begin{table}
  \begin{center}
    \caption{
      X-ray and radio detection rates in each class normalized to that of B+O+.
      Objects that could be contaminated by non-AGN X-ray/radio emission are
      excluded from the statistics (see text for details).
      Due to this conservative analysis, the numbers here should not
      be regarded as purity of the AGN sample.  The numbers also suffer
      from X-ray/radio sensitivity limits and the fractions are naturally
      lower in lower luminosity AGNs.
    }
    \label{tab:xray_freq}
    \begin{tabular}{r|ccc} 
             & X-ray & Radio\\\hline$
      O+B+$ & $1.000\pm0.186$ & $1.000\pm0.035$\\
      $O+B-$ & $0.097\pm0.097$ & $0.240\pm0.028$\\
      $O-B+$ & $0.242\pm0.108$ & $0.088\pm0.012$\\
      $O-B-$ & $0.015\pm0.011$ & $0.009\pm0.001$\\
      $O+Bn$ & $0.140\pm0.042$ & $0.258\pm0.010$\\
      $O-Bn$ & $0.012\pm0.012$ & $0.027\pm0.003$\\
      $OnBn$ & $0.040\pm0.020$ & $0.083\pm0.005$\\
    \end{tabular}
  \end{center}
\end{table}

\begin{figure*}
  \begin{center}
    \FigureFile(160mm,80mm){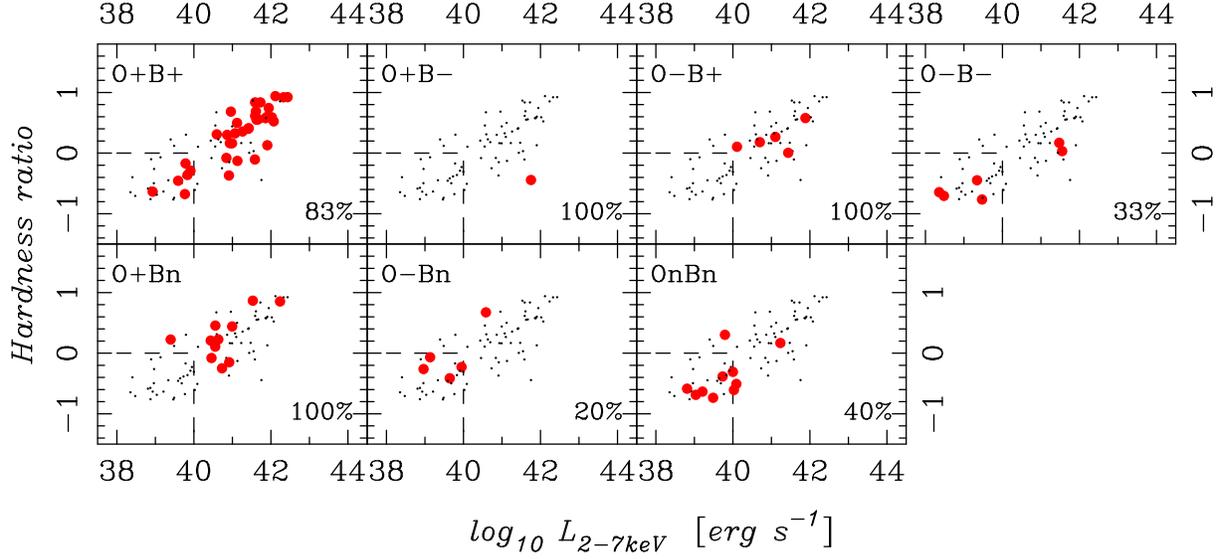}
  \end{center}
  \caption{
    Hardness ratio between soft ($0.5-1.2$ keV) and hard ($2.0-7.0$ keV) bands
    plotted against hard X-ray luminosity.
    The panels show subclasses of AGNs defined in Table \ref{tab:agn_class}.
    The dots are all the sources and the points are those in each class.
    The dashed rectangle on the bottom-left of each panel defines the 
    region, in which we expect contamination of star forming galaxies
    or other non-AGN sources.  The numbers in each panel shows the fraction of
    objects outside of the rectangle.
  }
  \label{fig:chandra_obj}
\end{figure*}

Due to the serendipitous nature of the X-ray catalog,
we can compare the relative frequency of X-ray detections
of various classes of objects defined in Table \ref{tab:agn_class}.
We normalize the X-ray detection frequency of the B+O+ objects
to unity and show the relative frequencies in Table \ref{tab:xray_freq}.
We also show the X-ray properties of the objects in Fig. \ref{fig:chandra_obj}.
There are several energy sources of X-ray emission: AGNs, supernova remnants,
high/low-mass X-ray binaries, and hot thermal plasma in the halos.
To quantify the dominance of AGNs in each panel, we define a region of the diagram
where we expect contamination of non-AGN sources as the dashed line.
This definition is motivated by the fact that these non-AGN
sources are likely soft, low-luminosity sources.  For example,
\citet{irwin03} showed that most of the low-mass X-ray binaries
have a hardness ratio less than 0 between the soft (0.3--1.0keV) and hard (2--6keV) bands.
The definition is further motivated to include most of the O-B- galaxies whose
X-rays are likely star formation or other non-AGN origins.
But, this is still a somewhat arbitrary definition and should not be
over-interpreted. 
AGNs may well populate in the dashed box.
Note that we have excluded these possible non-AGNs from the statistics
in Table \ref{tab:xray_freq} to be conservative.

We discuss each class of objects in what follows.

\begin{itemize}

\item {\bf O+B+ : } Most of the objects ($\sim80\%$) show
$L_X>10^{41}\rm \ erg\ s^{-1}$ with relatively hard spectra.
They are likely AGNs.

\item {\bf O+B- : }
We have only one X-ray source, which is likely an AGN due to
its high luminosity ($L_X\sim10^{42}\rm \ erg\ s^{-1}$).
We do not further discuss this class of objects due to the poor statistics.

\item {\bf O-B+ : }
Many of the objects in this class are relatively luminous X-ray sources
with medium hardness.  They are likely AGNs.
Our method misses this class of objects.

\item {\bf O-B- : } These objects are likely star forming galaxies.
Most of them have soft, low X-ray luminosity.  But, there are
luminous, medium hardness sources, which are likely AGNs.
The X-ray detection rate is only 2\% of that of O+B+
and such AGNs are very rare.

\item {\bf O+Bn : } These are AGN candidates identified by the Oxygen-excess
method, but their emission lines are too weak to apply the BPT diagnostics.
The distribution of objects in Fig. \ref{fig:chandra_obj} is relatively
similar to that of B+O+, and most objects in this class are likely AGNs.
The X-ray detection rate is not as high as O+B+, but it is probably because
AGNs in this class are weaker given the weak emission lines.
X-rays clearly show that
the Oxygen-excess method works well in identifying such weak AGNs.

\item {\bf O-Bn : } We do not observe any significant Oxygen
flux excess in those objects, and most of the X-ray sources
in this class are indeed low-luminosity soft sources with
an exception of a very hard source.
As inferred from the stacked spectrum, a small fraction of objects
in this class may be real AGNs.
But, we note a very small X-ray detection rate in Table \ref{tab:xray_freq} (1\%).

\item {\bf OnBn : }  X-ray detections of these objects may be surprising
as we do not observe any significant emission lines.
There seems a sequence of soft X-ray objects with X-ray luminosities
between $\sim10^{39}$ and $\sim10^{40}\rm \ erg\ s^{-1}$.
We find a clear correlation between
the X-ray luminosities of these objects and their stellar mass, which
lends support to the low-mass X-ray binary origin \citep{kim04}.
It may also be that diffuse thermal emission contributes to
the observed X-ray luminosity \citep{flohic06}.
There are a few sources with hardness ratio around 0.
We exclude the possibility of the supernova and high-mass X-ray binary origin
for these sources
because the host galaxies are quiescent galaxies with very little
on-going star formation.  Although we cannot be conclusive, their hardness ratios
seem to suggest that they are
unlikely due to low-mass X-ray binaries.
They may possibly be obscured AGNs with very weak optical emission lines.

\end{itemize}

The comparisons with X-ray sources give another quantitative estimate
of the robustness of the Oxygen-excess method.  Although X-rays
suffer from contamination from non-AGN sources, the numbers
from conservative analyses in Table \ref{tab:xray_freq}
and Fig. \ref{fig:chandra_obj}
suggest that the Oxygen-excess method
separates AGNs from star forming galaxies well,
although it does miss a fraction of  AGNs (e.g., O-B+).
The SF/AGN classifications of low-luminosity objects
for which BPT is not applicable (i.e., O+Bn and O-B-) are good (Fig. \ref{fig:chandra_obj}),
demonstrating its sensitivity to low-luminosity AGNs.

\subsection{FIRST sources}

We turn our attention to radio sources.
Radio wavelengths have also been used to identify (distant) AGNs (\cite{miley08}
and references therein).
In this subsection, we use data from the FIRST survey \citep{becker95,white97}
to study the properties of the Oxygen-excess objects.

The SDSS objects and FIRST sources are cross-matched within 1 arcsec.
We have confirmed that our results are not sensitive to the matching
radius (a matching radius of 2 arcsec gives the same results).
This is simple positional matching and we may well miss extended radio
sources such as jet lobes. However, such extended radio emissions are
relatively rare ($\sim10\%$; \cite{lin10}) and they should not strongly
alter our conclusions.
We do not reject extended sources
from the matching because AGN point sources may be buried under extended
radio emission due to star formation due to the poor spatial resolution of
FIRST.
In total, we have 1,747 matches in our sample.
We adopt $f_{integrated}$ from the FIRST catalog as radio power.
Note that our results remain essentially the same if we use $f_{peak}$.
We apply the $k$-correction to the radio power assuming a power-law
spectrum of $f_\nu\propto\nu^{-0.8}$ \citep{condon92}.

\begin{figure}
  \begin{center}
    \FigureFile(80mm,80mm){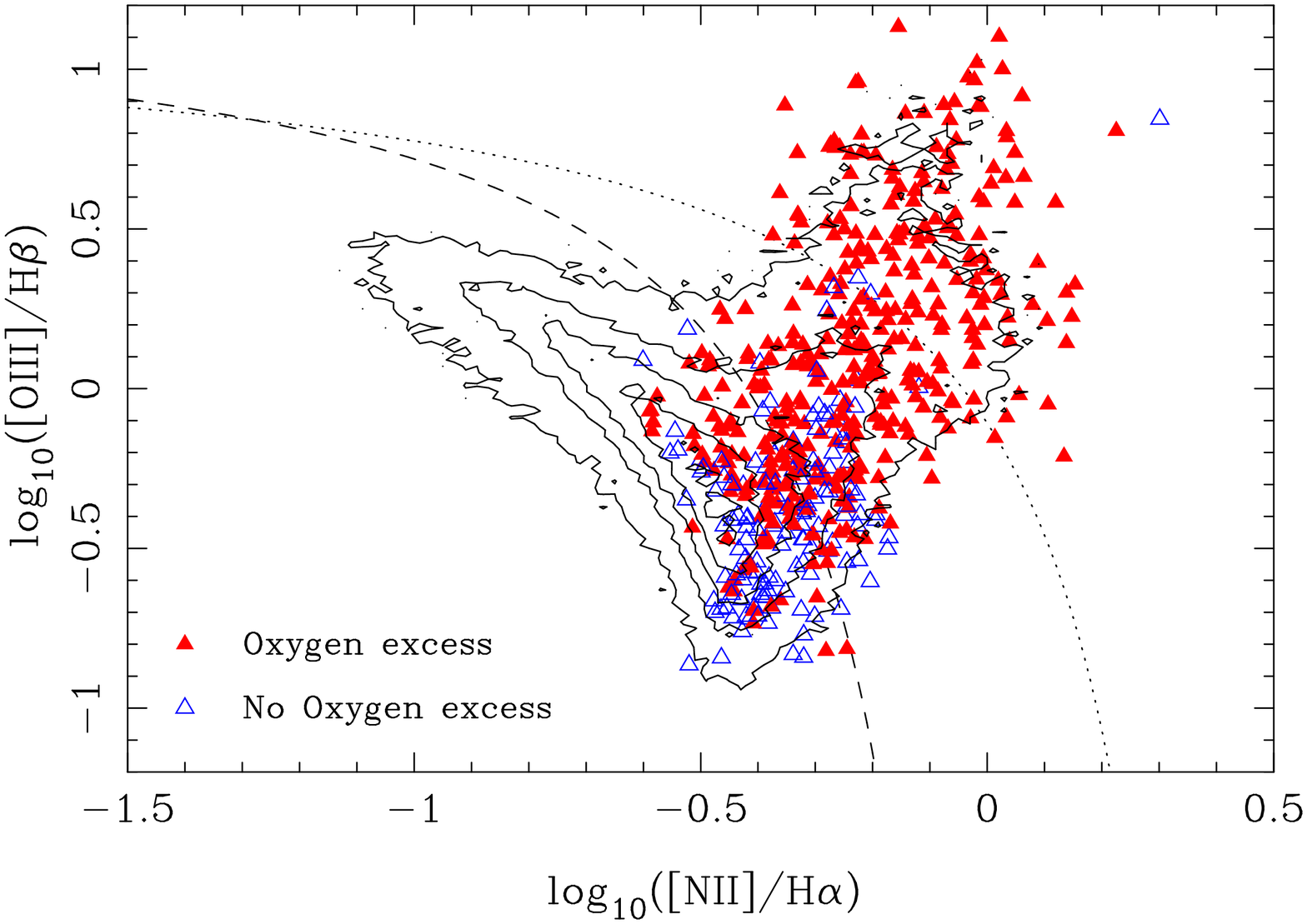}\hspace{0.5cm}
  \end{center}
  \caption{
    As in Fig. \ref{fig:bpt_xray}, but here we plot objects detected in the FIRST survey.
  }
  \label{fig:bpt_radio}
\end{figure}

We plot in Fig. \ref{fig:bpt_radio} the distribution of radio detected sources
on the BPT diagram.  
The radio detected objects tend to spread around the AGN sequence and
extend to the bottom of the SF sequence.
This is is in contrast to X-ray sources, which are
mostly located away from the \citet{kauffmann03} threshold curve
as shown in Fig. \ref{fig:bpt_xray}.
This is partly due to increased contamination from star formation
activities in radio wavelengths than in X-rays.

\begin{figure*}
  \begin{center}
    \FigureFile(160mm,80mm){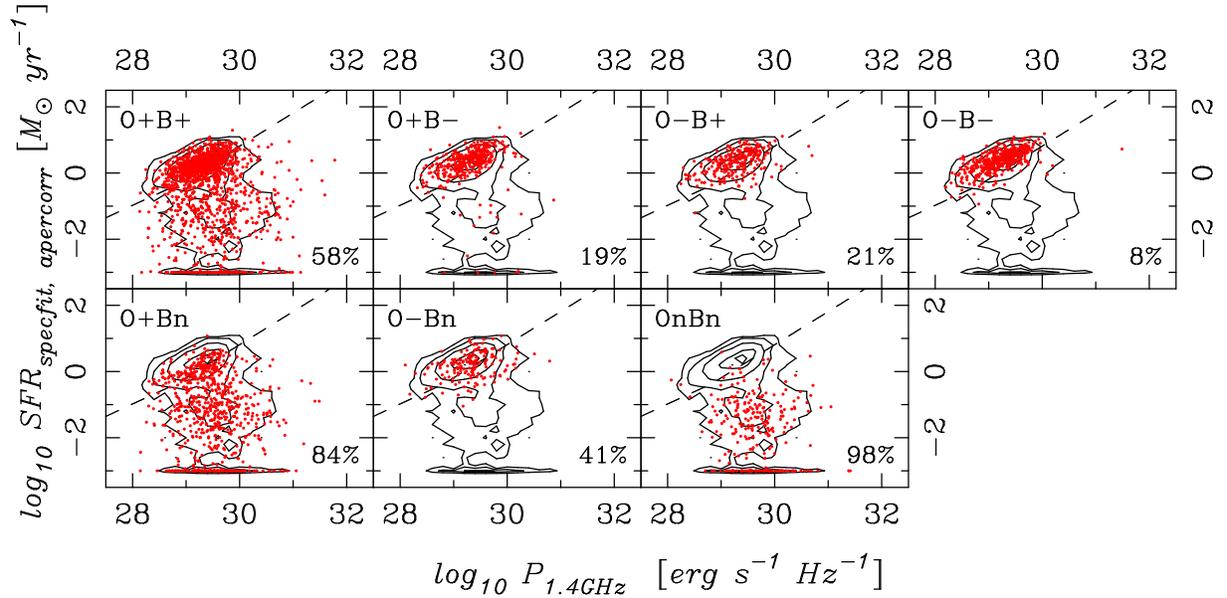}
  \end{center}
  \caption{
    SFRs against radio power.
    The panels show each class of AGNs defined in Table \ref{tab:agn_class}.
    The contours are all the radio sources and the dots are those in each class.
    The dashed curve is the SFR-radio power relation from \citet{hopkins03}
    shifted downwards by 0.7 dex (i.e., a factor of 5).
    Objects below the curve are likely dominated by AGNs and
    the numbers in each panel shows the fraction of objects below the curve.
  }
  \label{fig:first_obj2}
\end{figure*}

Fig. \ref{fig:first_obj2} shows SFRs from the spectral fits against
the radio power for each class of objects defined in Table \ref{tab:agn_class}.
A star formation sequence of galaxies can be seen at the top-left corner
of the diagram.  We shift the radio power -- SFR relation from \citet{hopkins03}
downwards by a factor of 5 to separate AGN-dominated and SF-dominated objects
as shown by the dashed curve.  A large fraction of O+B+ objects are dominated by
star formation, which is in contrast to X-rays (Fig. \ref{fig:chandra_obj}).
This illustrates the elevated contamination from star formation in radio
wavelengths.  But, we emphasize that not all the galaxies in the star forming
region of the diagram are pure star forming galaxies and they may well
host AGNs, whose radio 
emission due to AGN is weaker than that due to star formation.
We remove such galaxies from the analysis just to be conservative.

Despite the significant contamination from star formation,
the FIRST detection rates shown in Table \ref{tab:xray_freq}
are fairly useful.  
As expected, O+B+ and O-B- show the highest
and lowest radio detection rates, respectively.  Interestingly,
O+B- shows a relatively high detection rate.
This is encouraging because it shows that not all the objects
in this class are contamination and some of the O+B- objects are
real AGNs.
We comment on the other classes below.

\begin{itemize}
\item {\bf O+Bn : }
As expected from their weak emission lines, galaxies in this class do not
show strong on-going star formation.  Most of the objects are not in the
star forming region of the diagram and they are likely AGNs.
We note that those in the star forming region occupy only the bottom
half of the star forming sequence (compare with O-B-, for example),
suggesting that even these galaxies have some level of AGN activities.
We have shown from the stacking and X-ray analyses that this class of objects
are mostly real AGNs.  This radio analysis adds even further support to it.

\item {\bf O-Bn : }
Roughly 40\% of the radio detected sources
are likely AGNs.
If we compare the panels for O+Bn and O-Bn, the classifications look
reasonable, but it seems that we do missclassify a fraction of AGNs
in this class as already suggested in the stacking analysis.
However, the radio detection rate in Table \ref{tab:xray_freq}
is fairly low and we do not regard this as a big issue.

\item {\bf OnBn : }
Some of the apparently quiescent galaxies host AGNs.
The radio detection rate is 8\% of O+B+, which may be
relatively high for such quiescent galaxies.
This class of objects clearly shows that we cannot
identify all AGNs with optical methods only.
Deep multi-wavelength data are essential to sample the entire
AGN populations.

\end{itemize}

The radio detections provide another interesting test of the Oxygen-excess method.
While we do suffer from a level of contamination and incompleteness,
the Oxygen-excess method overall works well.  In fact, the radio
detection rates in Table \ref{tab:xray_freq} are the highest
where we observe excess Oxygen luminosities.

\vspace{0.3cm}
To summarize,
we have compared the Oxygen-excess method with BPT, X-ray and radio.
As mentioned at the begging of this section, it is hard to quantify
the purity and incompleteness of the Oxygen-excess method because
no AGN identification method gives a complete sampling of AGNs.  But, the stacked spectra
show that the SF/AGN classifications are on average very reasonable
and the X-ray and radio analyses give a support to it.
The results from these analyses all lead us to conclude that the Oxygen-excess method is
a good statistical tool to identify AGNs.  In particular, it is fairly
sensitive to low-luminosity AGNs for which BPT is not applicable.
It of course suffers from incompleteness (e.g., O-B+) and contamination
(e.g., a fraction of O+B- galaxies) as all the other methods do, but it classifies AGNs
and star forming galaxies well.  It is a powerful statistical
tool to identify AGNs from spectroscopic data.

\subsection{Ionizing source ---  accreting material or evolved stars?}

Finally, we discuss contamination of objects that are photo-ionized by non-AGN sources.
Most of the identified AGNs are low-luminosity, low-ionization objects
(LINERs: \cite{heckman80}) as quantified
in Paper-II and there are several possible origins of LINER-like spectra:
low-ionization AGNs \citep{ho93,ho97},
shocks due to supernova/jets \citep{cox72,heckman80,dopita95,dopita96},
Wolf-Rayet stars \citep{terlevich85,kewley01},
post-starburst \citep{taniguchi00}, and
post-AGB stars \citep{binette94,stasinska08,sarzi10,cidfernandes11}.
LINERs likely constitute a heterogeneous class of objects \citep{ho08}.
The existence of broad permitted lines in a fraction of LINERs
provides strong evidence for the AGN origin \citep{ho93,ho97},
but relative contributions of the other ionizing mechanisms
to the overall LINER population remain unclear.

Let us focus on the O+Bn class, in which more than $60\%$ of the
Oxygen-excess objects fall.  These objects show only weak emission lines
and therefore their underlying star formation activities are weak
(otherwise we would have observed strong emission lines).
For such objects, we can reject Wolf-Rayet stars and supernova
as a primary cause
of the Oxygen flux excess because these sources play a role
only in star forming galaxies.
The post-starburst origin is unlikely to produce such a large number
of AGNs and unlikely a primary cause, too.
\citet{cidfernandes11} suggest that galaxies with EW(H$\alpha)<3\rm\AA$
are likely due to post-AGB stars based on a photo-ionization calculation
of a single burst stellar population.  
We find that a significant fraction (75\%) of
O+Bn objects have EW(H$\alpha)<3\rm\AA$.
But, we argue that many of them are actually AGNs.
We find that two-thirds of O+Bn objects detected in X-rays (excluding
those in the dashed region of Fig. \ref{fig:chandra_obj})
have EW(H$\alpha)<3\rm\AA$.
We obtain a fairly similar number for those detected in radio (67\%;
again, those dominated by star formation are excluded from the statistics).
Recently, \citet{capetti11} also reported on a radio detection of quiescent
galaxies in SDSS.
These X-ray and radio detected sources are probably real AGNs and
this suggests that EW(H$\alpha)<3\rm\AA$ does not necessarily mean
that the observed Oxygen-excess is due to post-AGB stars.
The fact that a fraction of OnBn objects are detected
in radios (Table \ref{tab:xray_freq}) adds a further line of argument
against EW(H$\alpha$)$<3\rm\AA$.

However, recent integral spectroscopy of nearby galaxies has revealed 
diffuse extended emission in early-type galaxies and seem to provide
evidence of contributions from post-AGB stars.
\citet{sarzi10} observed that nearby early-type galaxies with
radio detections show extended emission.
These galaxies show LINER-like emission line ratios and \citet{sarzi10}
suggested that post-AGB stars can supply enough ionizing photons
to explain the observation and thus the photo-ionization is primarily due to
post-AGB stars.  However, an uncertainty in the fraction
of ionizing photons that are reprocessed into emission and our limited
understanding of the last stage of the stellar evolution seem to hamper a firm conclusion.
Extended line emission is also recently observed by \citet{yan11b}.
They claimed that the ionizing parameter increases towards larger radii
from the galaxy center and the most natural explanation of it would be
due to post-AGB stars, although they found that post-AGB stars cannot
supply enough ionizing photons.
One can turn the question around and ask whether AGN can supply enough
photons to explain the observed emission lines.
\citet{maoz98} and \citet{eracleous10b} reported that a fraction of
LINERs show a severe deficit of ionizing photons and need other
ionization sources.
From these observations, there is no doubt that the Oxygen-excess
objects are contaminated by the non-AGN emission and at least a fraction of
observed emission line luminosities is likely due to post-AGB stars.

However, the contributions from post-AGB stars may not be very significant.
We will discuss in depth in Section 3.3 of Paper-II,
but we briefly outline our argument here.
We observe a clear correlation between optical emission line luminosities
and hard X-ray luminosities (Fig 3 of Paper-II).  Post-AGB stars are not
luminous in hard X-rays and the hard X-ray luminosity is a good measure of
AGN activity.  On the other hand, optical emission lines can be
significantly contaminated by post-AGB photo-ionization.
The observed clear correlation suggests that
the contribution from post-AGB stars to the observed optical emission
is not severe.  The hard X-ray detected sources have typical properties
of the Oxygen-excess objects (they have stellar mass of $>10^{10}$
and are mostly red galaxies with low SFRs of $<0.1\rm M_\odot\ yr^{-1}$),
thus they represent the Oxygen-excess objects well.
One might worry that non-AGN sources such as low-mass X-ray binaries
may be contributing to the observed X-rays because these binaries often
significantly contribute to the overall X-ray emission in massive
quiescent galaxies.  But, we show in Fig. 5 of Paper-II that
the hard X-ray luminosity does not correlate with stellar mass.
The rather weak dependence of X-ray luminosity on the host galaxy mass
is also observed by other authors \citep{mullaney11,aird11}.
This excludes the low-mass X-ray binary origin of the observed X-rays
because their contribution should increase with increasing stellar mass
\citep{kim04}.
The most likely origin of the hard X-ray is therefore AGN and the clear correlation
between hard X-ray and optical luminosity shown in Paper-II suggests that
the contamination from post-AGB stars to the observed emission line
is not significant.

The X-ray sample is a just small portion of the entire Oxygen-excess objects.
We make another subsample of them to further quantify the role of post-AGB stars.
This is a subsample of quiescent (i.e., SFR is zero) Oxygen-excess objects
in a narrow redshift slice to eliminate any redshift effects.
The emission due to post-AGB stars should correlate strongly with
stellar mass contained within the area covered by the fibers,
while the AGN emission is unlikely to be strongly correlated
with mass \citep{aird11,mullaney11}.
Within the narrow redshift slice, we find that observed emission line
luminosities only weakly depend on stellar mass, which suggests 
that post-AGB stars do not significantly contribute to the overall emission.
Based on a very simple model, we find that 23\% of the emission is due
to post-AGB stars in typical Oxygen-excess objects with stellar mass of
$\rm10^{10}\ M_\odot$ within the fibers.

It seems that a large fraction of O+Bn objects are likely AGNs.
But, we still do not know the exact abundance of non-AGNs
and the exact fractional contribution of the post-AGB photo-ionization to
the observed emission line luminosities.
We may well have Oxygen-excess objects whose emission is completely
powered by post-AGB stars.
It has been a challenging task to pin down the abundance of true/false AGNs
\citep{ho08} and this would probably require deep multi-wavelength observations
of the nuclear region of well defined sample of galaxies.
The SDSS fibers subtend 3 arcsec on the sky and they include
a substantial fraction of bulge and disk components under
the typical seeing conditions of the site ($\sim1.5$ arcsec).
We deem that the SDSS data are not suited to address the issue in depth.
Also, our poor understanding of the last phase of the stellar evolution
puts further limit on our ability to constrain the role of post-AGB stars.
For these reasons, we do not try to go further from here.
This unknown fraction of contaminating non-AGNs (although it will be small)
remains one of the major uncertainties in results presented in Paper-II
and a more detailed study of both observational and theoretical aspects
would be needed to put a more stringent constraint on the role of
non-AGN photo-ionization.

\section{Summary}

We have developed a novel technique to identify AGNs based on 
a very simple idea of comparing expected and observed emission line luminosities.
We perform a spectral fits of the SDSS galaxies to obtain
SFRs and dust extinction, from which we can compute
expected emission line luminosities due to star formation.
By comparing the expected luminosities with observed luminosities,
we can statistically identify AGNs.
In this comparison, we use {\sc [oii]+[oiii]} luminosities.
This choice is motivated by the fact that AGNs span a wide range
in ionization state as quantified in Paper-II.

To test the newly developed method, we have made extensive
comparisons with the other AGN identification methods,
namely, BPT, X-rays and radio.
Our method suffers from contamination and incompleteness
as all the other methods do. 
But, the average properties from the stacked spectra and
the detection rates of X-ray and radio sources all suggest
that the Oxygen-excess method is a good statistical method
to identify AGNs.
The most unique feature of the method is its sensitivity.
We have demonstrated that our method is applicable to $\sim80\%$
of the galaxies, while BPT can be applied to only $\sim40\%$.
All the analyses above show that
the Oxygen-excess method works fairly well in identifying such low-luminosity
objects.
Another very unique feature, which we have not emphasized in this paper,
is its capability to subtract star formation
component from the observed emission line luminosity to extract
pure AGN emission, which is crucial to characterize AGN activities.
We will make an extensive use of these features
to study the nature of low-luminosity AGNs in Paper-II.

To summarize, the strengths and weaknesses of our method would be:

\noindent
{\bf STRENGTHS:}
\begin{itemize}
\item It requires only a sum of {\sc [oii]} and {\sc [oiii]}.
Note that the BPT diagnostics requires 4 lines
and it takes ratios of the lines, meaning that one needs to
detect each line at a sufficiently significant level.
H$\beta$ is often the weakest line in AGNs and that limits
the sensitivity of BPT.

\item It does not require H$\alpha$, allowing us to go up to $z\sim1$
with optical spectrographs.  If one uses {\sc [oii]} only at
a risk of missing high ionization AGNs, one can go up to even
higher redshifts of $z\sim1.7$.

\item It is fairly sensitive to low-luminosity AGNs that cannot be
identified by the BPT diagnostics.

\item It is able to subtract emission line fluxes due to
star formation and extract AGN fluxes.  We will make a full use
of this feature in Paper-II.

\end{itemize}

\noindent
{\bf WEAKNESSES:}
\begin{itemize}
\item It requires a robust continuum detection in well-calibrated spectra
(i.e., this method is applicable only to bright galaxies).
But, we deem that  multi-wavelength photometry could be used when continuum spectra
are not available.

\item It misses a fraction of AGNs selected from the BPT diagnostics,
X-rays, and radio.

\item It suffers from contamination of star forming galaxies.
But, as emphasized throughout this paper,
all methods suffer from incompleteness and contamination.

\item It misses weak AGNs in actively star forming galaxies as we will
fully quantify in Paper-II.

\end{itemize}

\vspace{0.5cm}
We would like to thank Masataka Fukugita and John Silverman for
extensive discussions, Luis Ho, Toru Nagao, and Yoshihiro Ueda,
for useful conversations,
Naoki Yasuda for providing the computing environment,
and Yen-Ting Lin for helpful comments on the paper.
This work was supported by World Premier International Research
Center Initiative (WPI Initiative), MEXT, Japan and in part by
KAKENHI No. 23740144.  This research has made use of data obtained from
the Chandra Source Catalog, provided by the Chandra X-ray Center (CXC)
as part of the Chandra Data Archive.
We thank the anonymous referee for his/her useful comments, which 
helped improve the paper.

Funding for the Sloan Digital Sky Survey (SDSS) and SDSS-II
has been provided by the Alfred P. Sloan Foundation, the
Participating Institutions, the National Science Foundation,
the U.S. Department of Energy, the National Aeronautics and
Space Administration, the Japanese Monbukagakusho, and
the Max Planck Society, and the Higher Education Funding
Council for England. The SDSS Web site is http://www.sdss.org/.

The SDSS is managed by the Astrophysical Research Consortium
(ARC) for the Participating Institutions. The Participating
Institutions are the American Museum of Natural History,
Astrophysical Institute Potsdam, University of Basel,
University of Cambridge, Case Western Reserve University,
The University of Chicago, Drexel University, Fermilab,
the Institute for Advanced Study, the Japan Participation Group,
The Johns Hopkins University, the Joint Institute for Nuclear
Astrophysics, the Kavli Institute for Particle Astrophysics
and Cosmology, the Korean Scientist Group, the Chinese Academy 
of Sciences (LAMOST), Los Alamos National Laboratory,
the Max-Planck-Institute for Astronomy (MPIA), the 
Max-Planck-Institute for Astrophysics (MPA), New Mexico
State University, Ohio State University, University of Pittsburgh,
University of Portsmouth, Princeton University, the United 
States Naval Observatory, and the University of Washington.

\appendix

\section{Extinction law and non-solar metallicity models}

In this appendix, we justify the choice of the \citet{cardelli89}
extinction curve over the \citet{charlot00} curve.
We also justify the exclusion of non-solar metallicity models.

\begin{figure*}
  \begin{center}
    \FigureFile(80mm,1mm){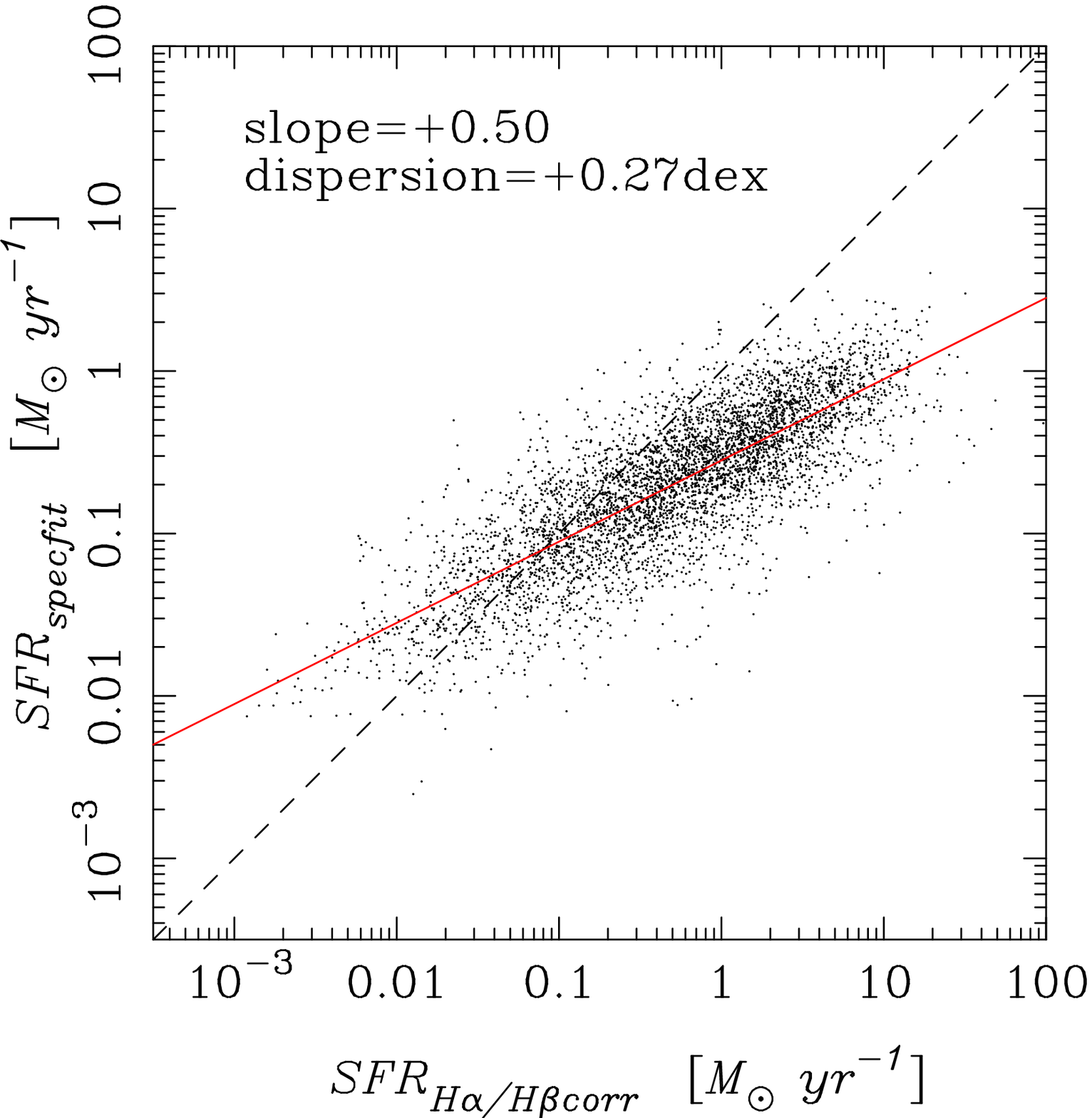}\hspace{0.5cm}
    \FigureFile(80mm,1mm){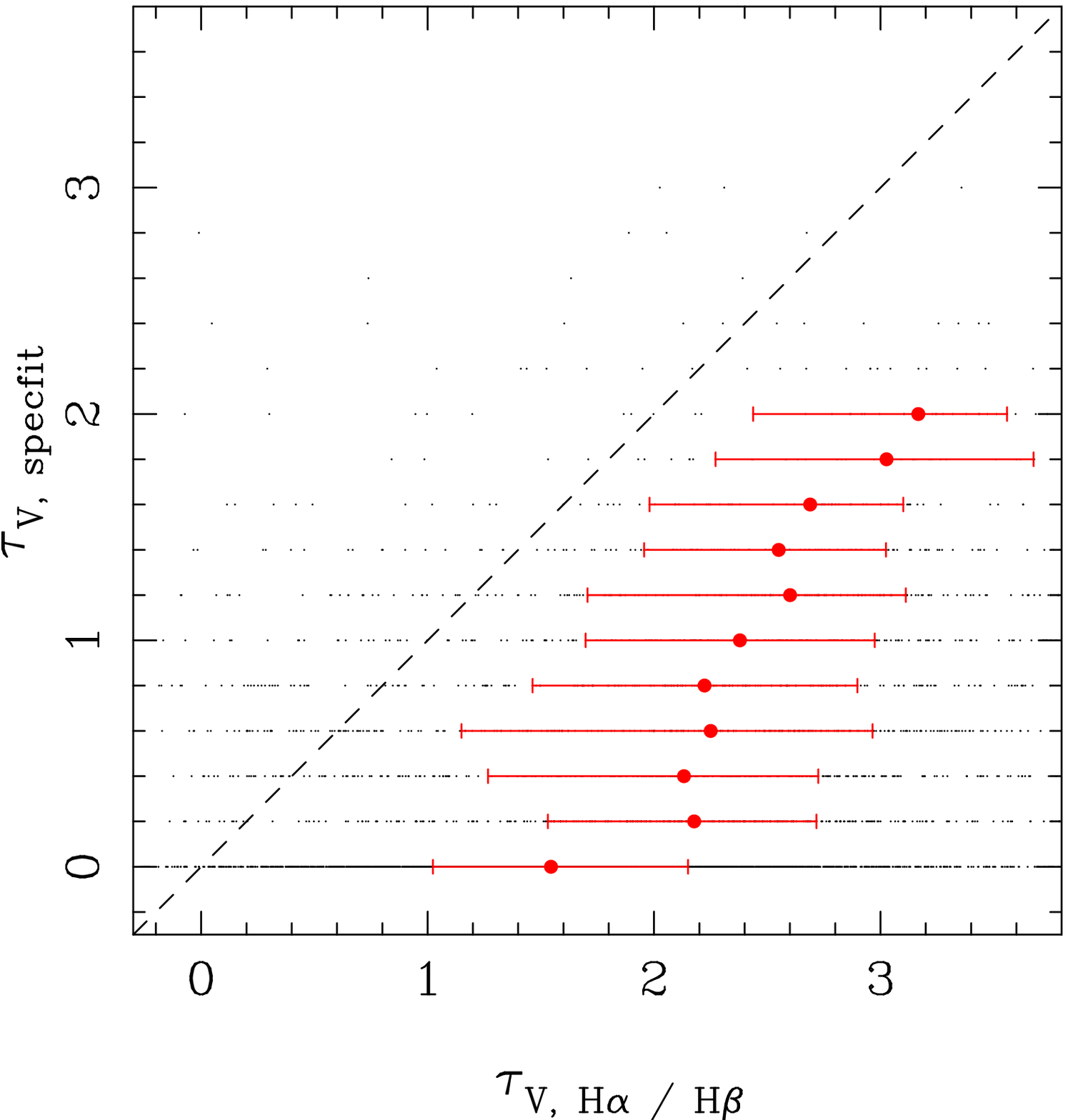}
  \end{center}
  \caption{
    As in Fig. \ref{fig:model_vs_obs}, but here we assume the \citet{charlot00}
    extinction curve.
    Plotted are a small subset ($\sim20,500$ objects) of the entire sample.
  }
  \label{fig:model1}
\end{figure*}

The reason why we prefer the \citet{cardelli89} curve is because
it gives better agreement between SFRs from the spectral fits
and those from H$\alpha$ corrected for the extinction.
As shown in the left plot of Fig. \ref{fig:model_vs_obs}, 
our SFR estimates are reasonably accurate, although there is a tilt.
This tilt becomes slightly larger if we adopt the \citet{charlot00} extinction
curve as shown in Fig. \ref{fig:model1}.
Also, the mean offset between SFRs from the spectral fits and
those from H$\alpha$ and H$\beta$ increases.
Furthermore, we find that the discrepancy between the dust extinctions from
the fits and those from the balmer decrement becomes larger
if we use \citet{charlot00}.
Because we extensively use SFRs and dust estimates from the spectral
fits, we prefer to use the \citet{cardelli89} extinction curve
to obtain better SFRs and dust estimates.
It might appear at odds to change only the extinction curve of
\citet{charlot00}, while keeping the two component dust model unchanged.
But, the two component model is physically sensible and
the observed better agreement with SFRs and dust justifies
the modification of the extinction curve.
We further note that \citet{charlot00} calibrated their model parameters
using starburst galaxies, but a very small portion of our sample is undergoing
such activities.
It is interesting note that we can predict emission line fluxes
equally well even if we adopt the \citet{charlot00} model.
It is likely due to the degeneracies between SFR and dust extinction

\begin{figure}
  \begin{center}
    \FigureFile(80mm,1mm){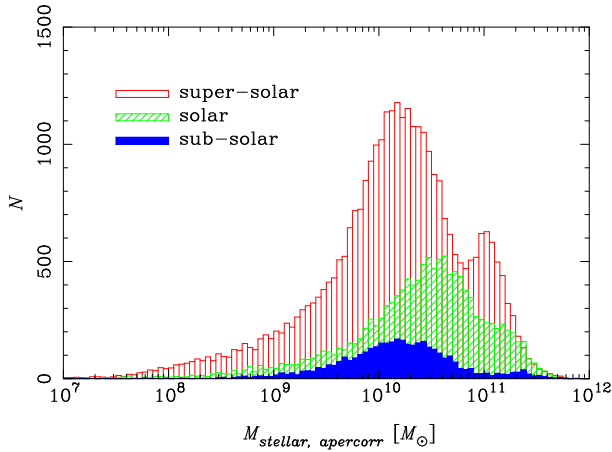}\hspace{0.5cm}
  \end{center}
  \caption{
    Stellar mass distributions.  The open, shaded, and filled histograms
    show galaxies fit with super-solar, solar,
    and sub-solar metallicity models, respectively.
  }
  \label{fig:metallicity_dep}
\end{figure}

Next, we turn to the metallicity dependence of our model fits.
We generate the model templates with super-solar, solar, and sub-solar
metallicities and perform the spectral fits using all the templates.
Fig. \ref{fig:metallicity_dep} shows the distribution of stellar mass
of galaxies.
Our model fits do not reproduce the mass-metallicity relation
(more massive galaxies are more metal-rich, e.g., \cite{nelan05}),
but the stellar mass distributions in Fig. \ref{fig:metallicity_dep}
do not show any strong dependence of stellar mass on metallicity.
Fig. \ref{fig:metallicity_dep} shows that we cannot estimate
metallicity of galaxies from our spectral fits.
This would not be too surprising because metallicity estimates
require careful absorption line diagnostics to break the age-metallicity
degeneracy \citep{worthey94}.
Furthermore, sub/super solar metallicity models degrade the accuracy
of our emission line flux estimates (a significant amount of
galaxies have H$\alpha$/H$\beta<2.86$) due to  inaccurate
corrections for stellar absorption.
For these reasons, we use only the solar metallicity models in
our spectral analyses.
We note that \citet{asari07} obtained a correlation between
stellar metallicity and gas-phase metallicity from versatile
spectral fits albeit with a significant scatter.

\section{Veilleux \& Osterbrock diagrams}

\citet{veilleux87} extended the commonly used BPT diagnostics and
showed that the {\sc [oi]} and {\sc [sii]} lines are also sensitive
to the presence of AGNs.  We have made extensive comparisons between
the Oxygen-excess and BPT diagnostics, but here we make further comparisons
with the Oxygen-excess and \citet{veilleux87} diagrams.

\begin{figure*}
  \begin{center}
    \FigureFile(80mm,80mm){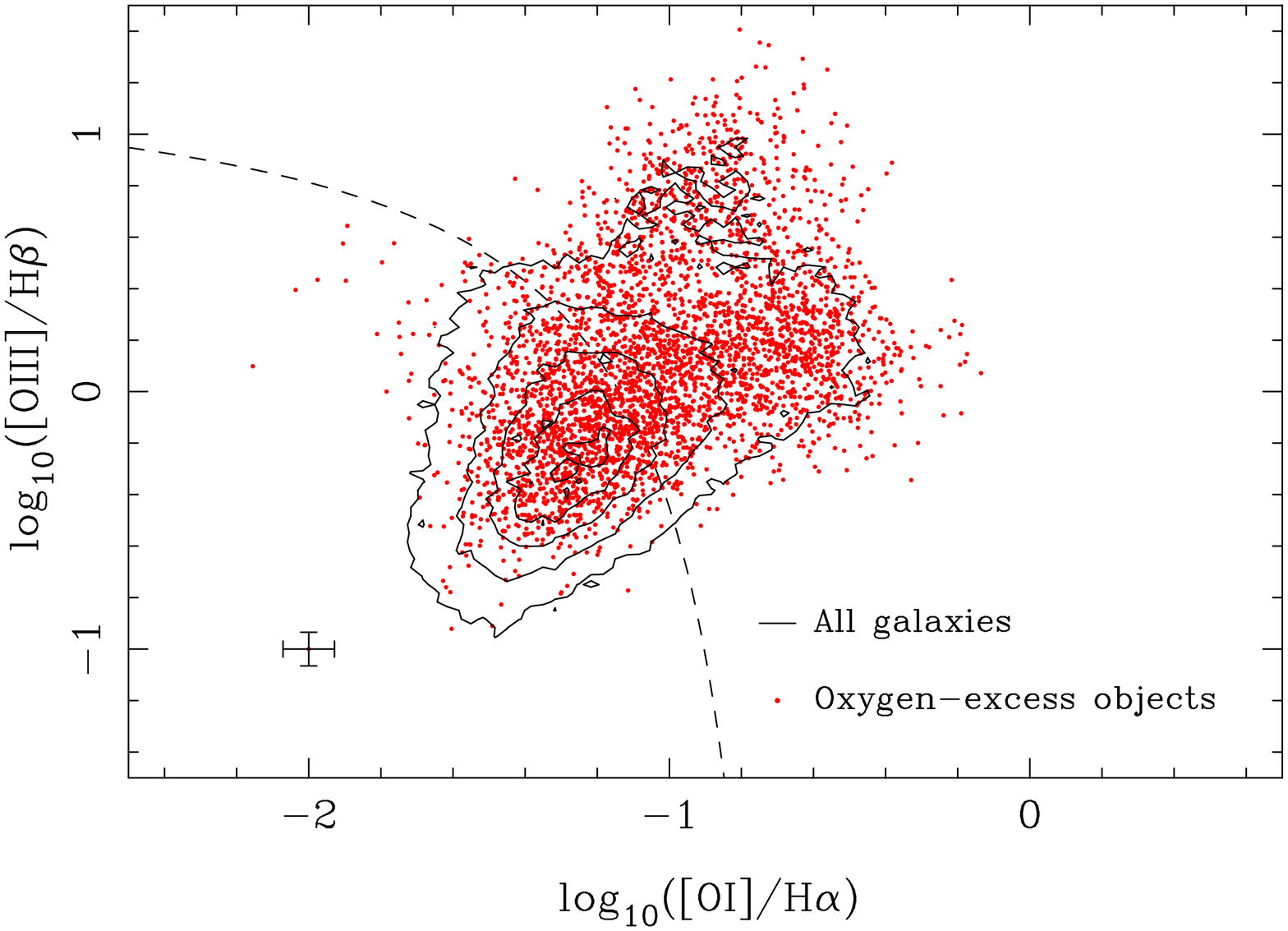}\hspace{0.5cm}
    \FigureFile(80mm,80mm){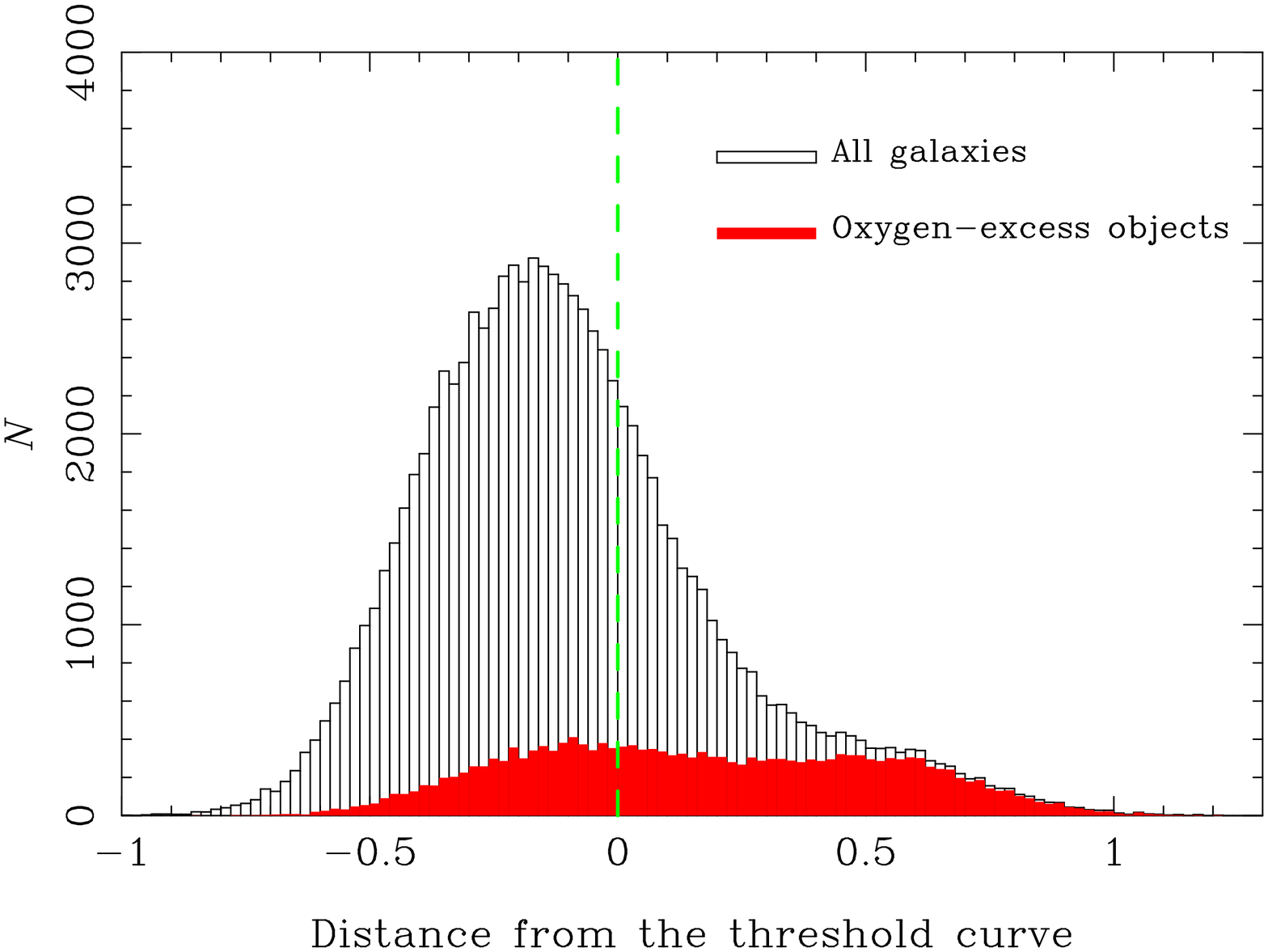}
  \end{center}
  \caption{
    Same as in Fig. \ref{fig:bpt_diagram}, but here we use {\sc [oi]}/H$\alpha$
    in place of {\sc [nii]}/H$\alpha$.  The dashed line to separate star forming
    galaxies from AGNs is from \citet{kewley01}.
  }
  \label{fig:bpt_diagram2}
\end{figure*}

\begin{figure*}
  \begin{center}
    \FigureFile(80mm,80mm){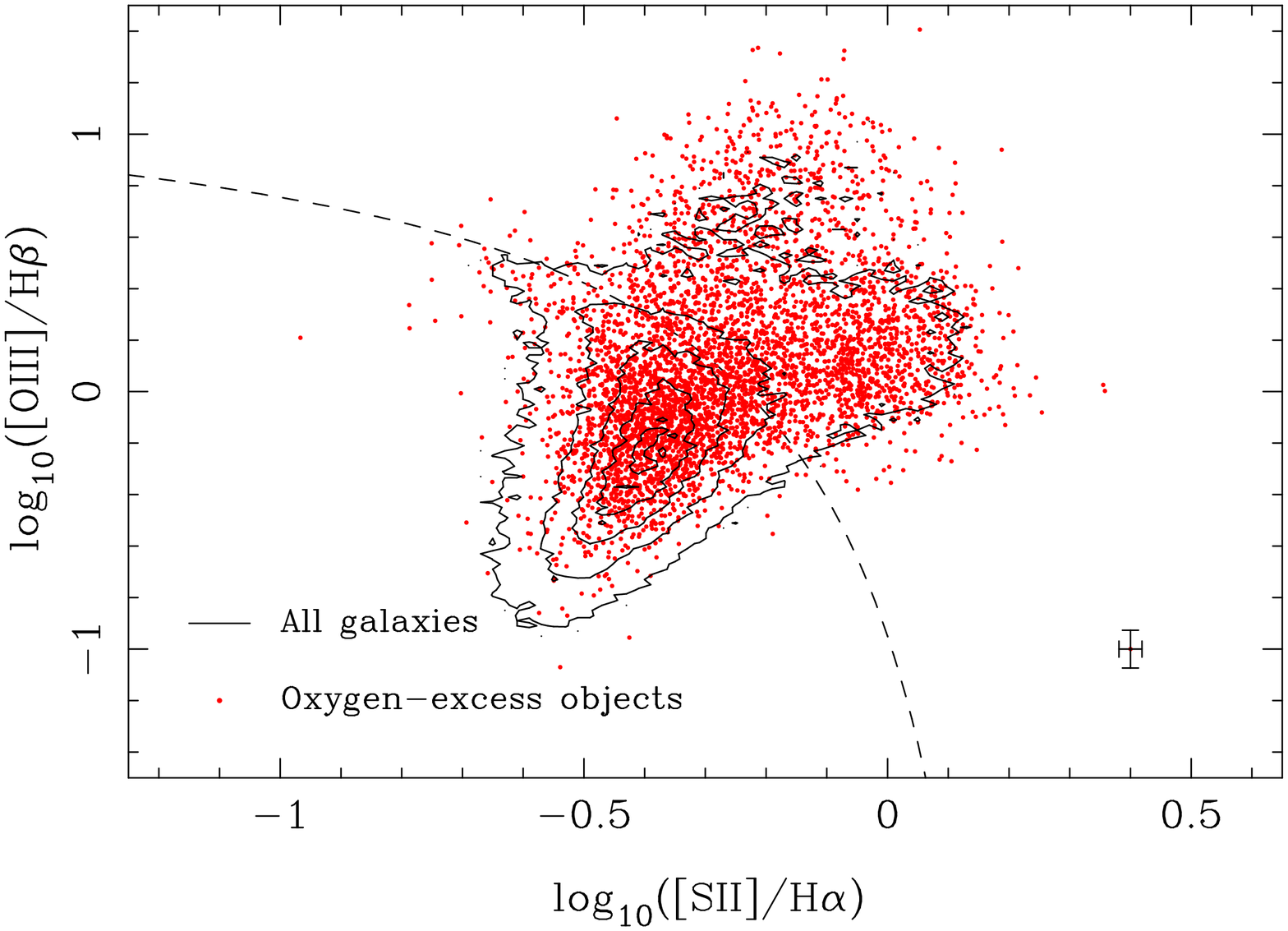}\hspace{0.5cm}
    \FigureFile(80mm,80mm){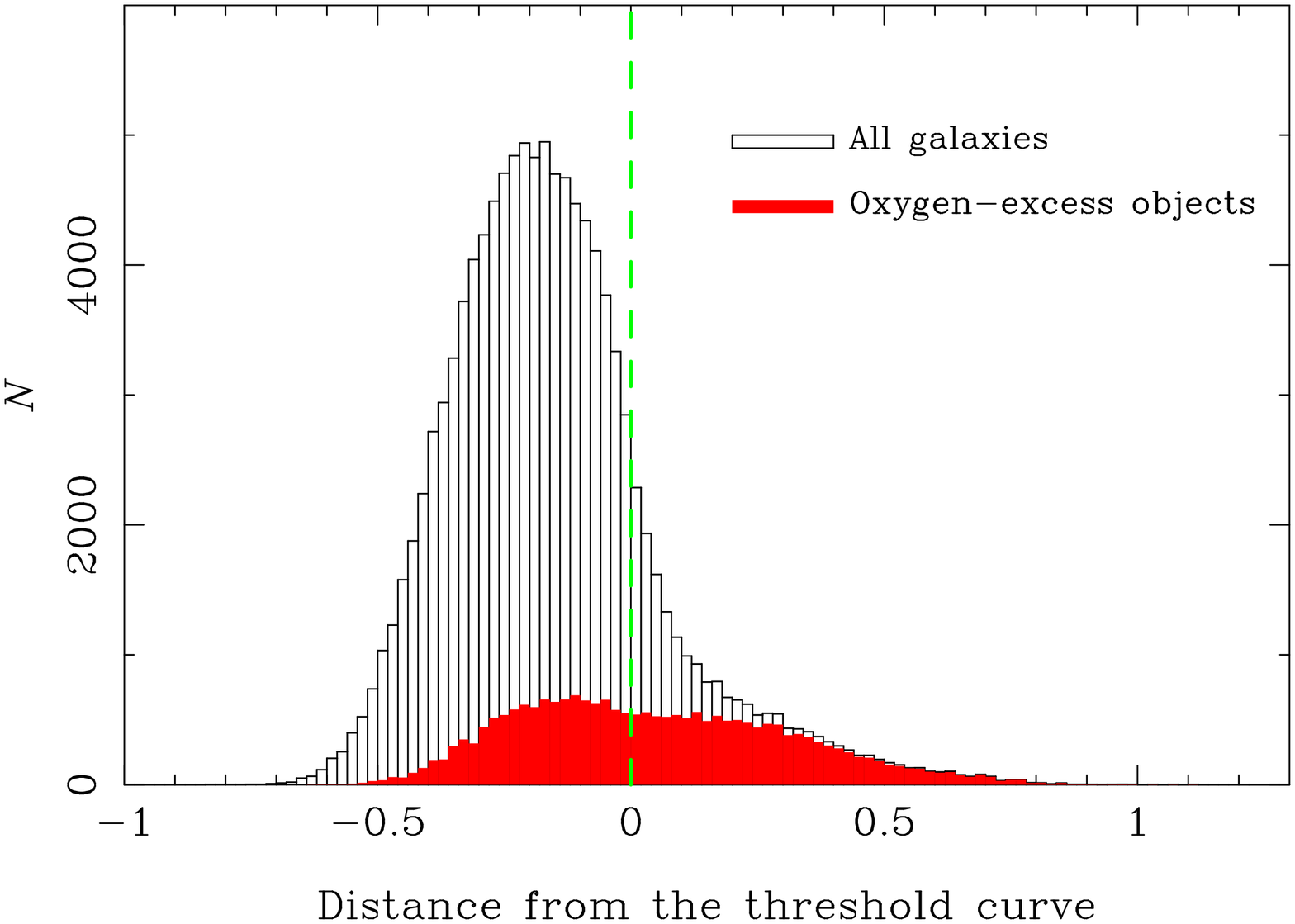}
  \end{center}
  \caption{
    Same as in Fig. \ref{fig:bpt_diagram}, but here we use {\sc [sii]}/H$\alpha$
    in place of {\sc [nii]}/H$\alpha$.  The dashed line to separate star forming
    galaxies from AGNs is from \citet{kewley01}.
  }
  \label{fig:bpt_diagram3}
\end{figure*}

We present the distributions of Oxygen-excess on the \citet{veilleux87}
diagrams in Figs. \ref{fig:bpt_diagram2} and \ref{fig:bpt_diagram3}.
The distributions of all galaxies do not show a clear branch of AGNs
like the one seen in Fig. \ref{fig:bpt_diagram}.  As mentioned in
the main body of the paper, the clear AGN sequence in the BPT diagram
is likely due to the secondary nature of nitrogen.
Theoretical modeling of the \citet{veilleux87} diagrams
shows a strong overlap between star forming galaxies and AGNs on
these diagrams \citep{stasinska06}.  In fact, the Oxygen-excess objects
spread over both the star forming and AGN regions of the diagrams.

\begin{figure*}
  \begin{center}
    \FigureFile(80mm,80mm){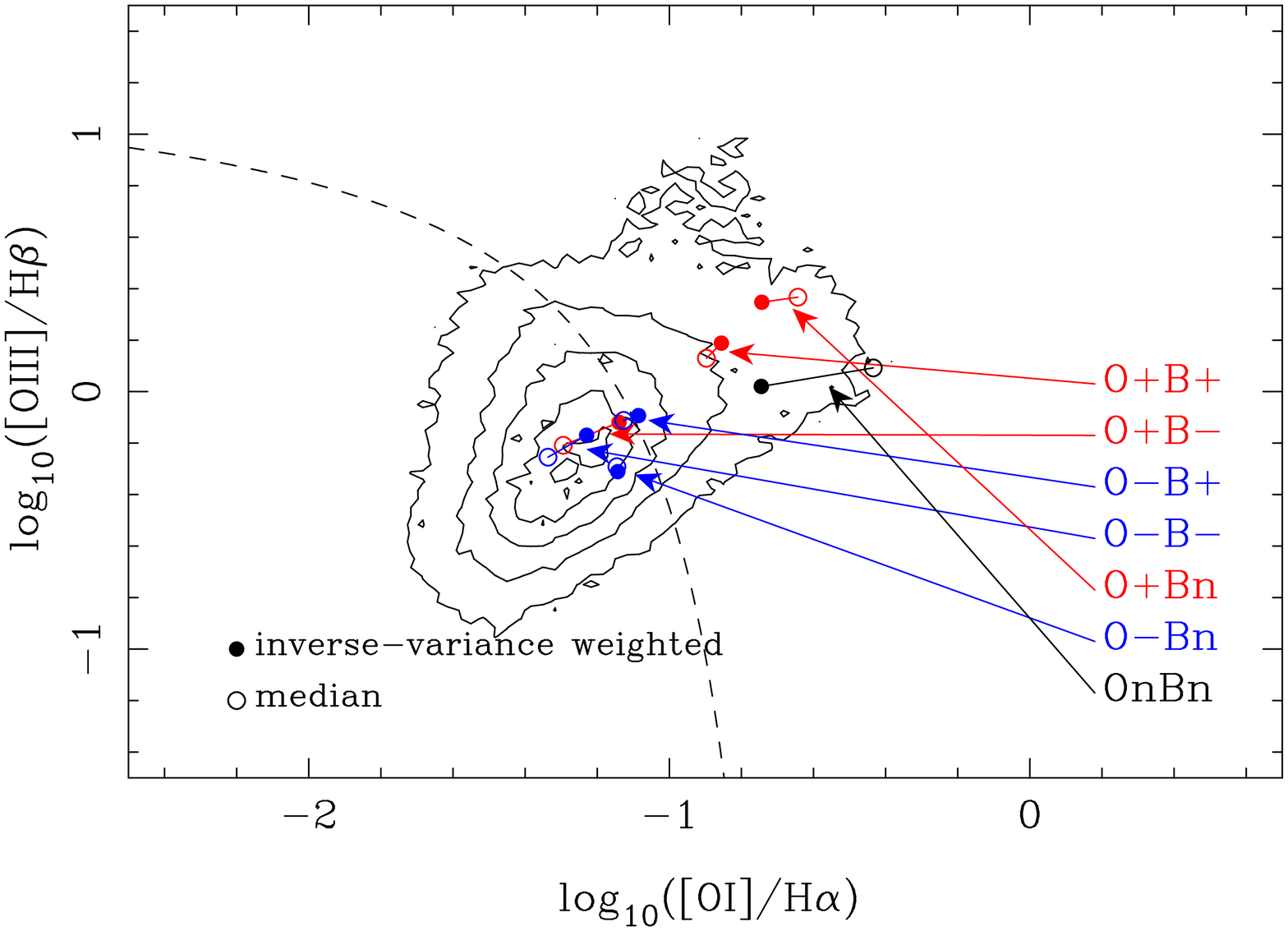}\hspace{0.5cm}
    \FigureFile(80mm,80mm){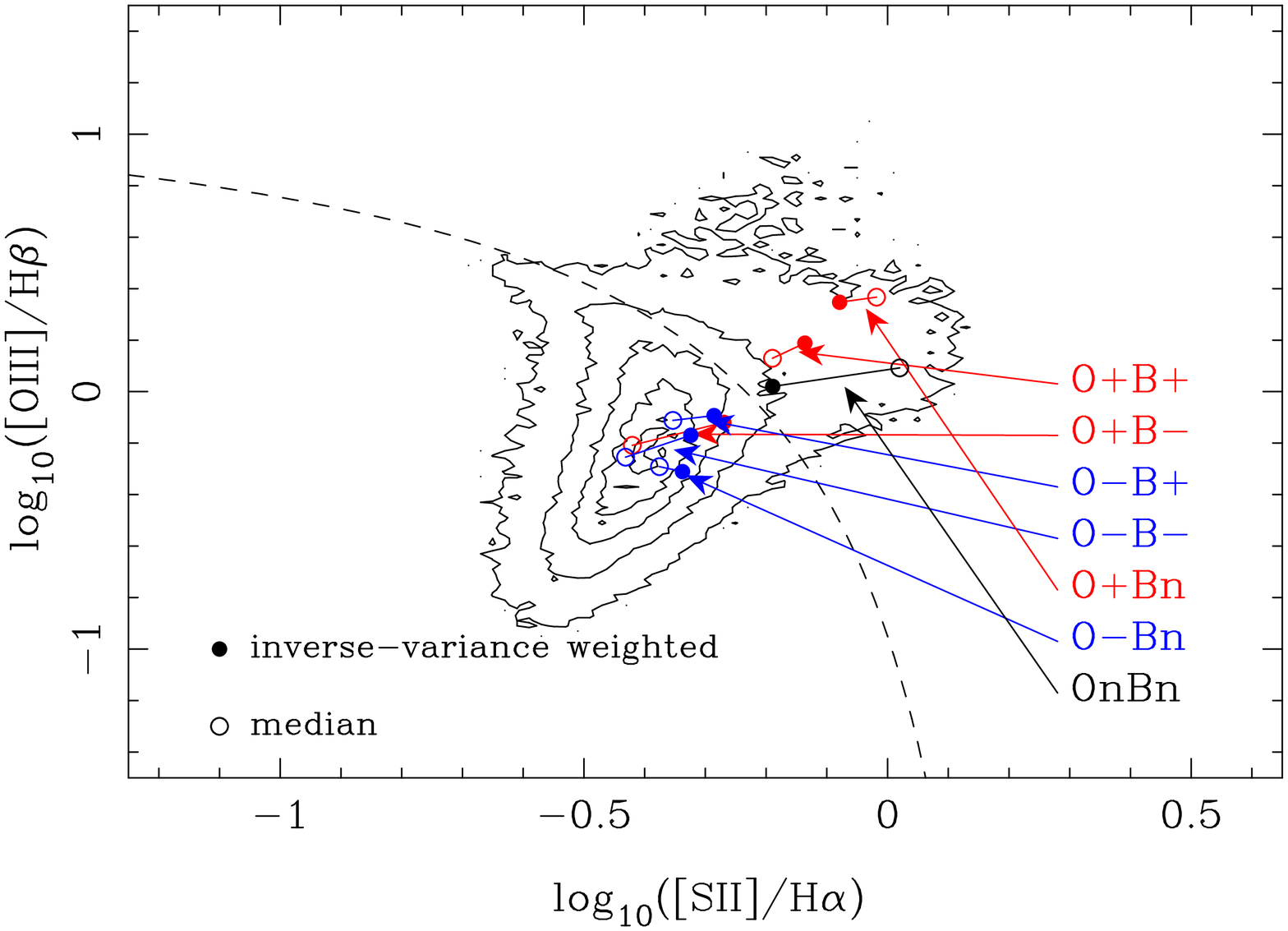}\\
  \end{center}
  \caption{
    Same as Fig. \ref{fig:bpt_diagram_stacked}, but for the two \citet{veilleux87} diagrams.
    The locations of the stacked objects are
    indicated by the points and arrows. The filled and open 
    circles are measured from the inverse-variance weighted stacking
    and from the median stacking, respectively.
  }
  \label{fig:bpt_diagram2_stacked}
\end{figure*}

Fig. \ref{fig:bpt_diagram2_stacked} shows the locations of the stacked
objects on the \citet{veilleux87} diagrams.  The trend is similar to
what observed in the BPT diagram (Fig. \ref{fig:bpt_diagram_stacked}),
albeit with a larger degeneracy between star forming galaxies and AGNs.
On these diagrams, only O+B+, O+Bn, and OnBn are clearly in the AGN
region.  The other classes are fairly close to each other and are all
in the star forming region.  But, as mentioned above, this does not
necessarily mean that they do not host AGNs because the star forming
sequence and AGN sequence overlap on these diagrams.

\begin{figure*}
  \begin{center}
    \FigureFile(80mm,80mm){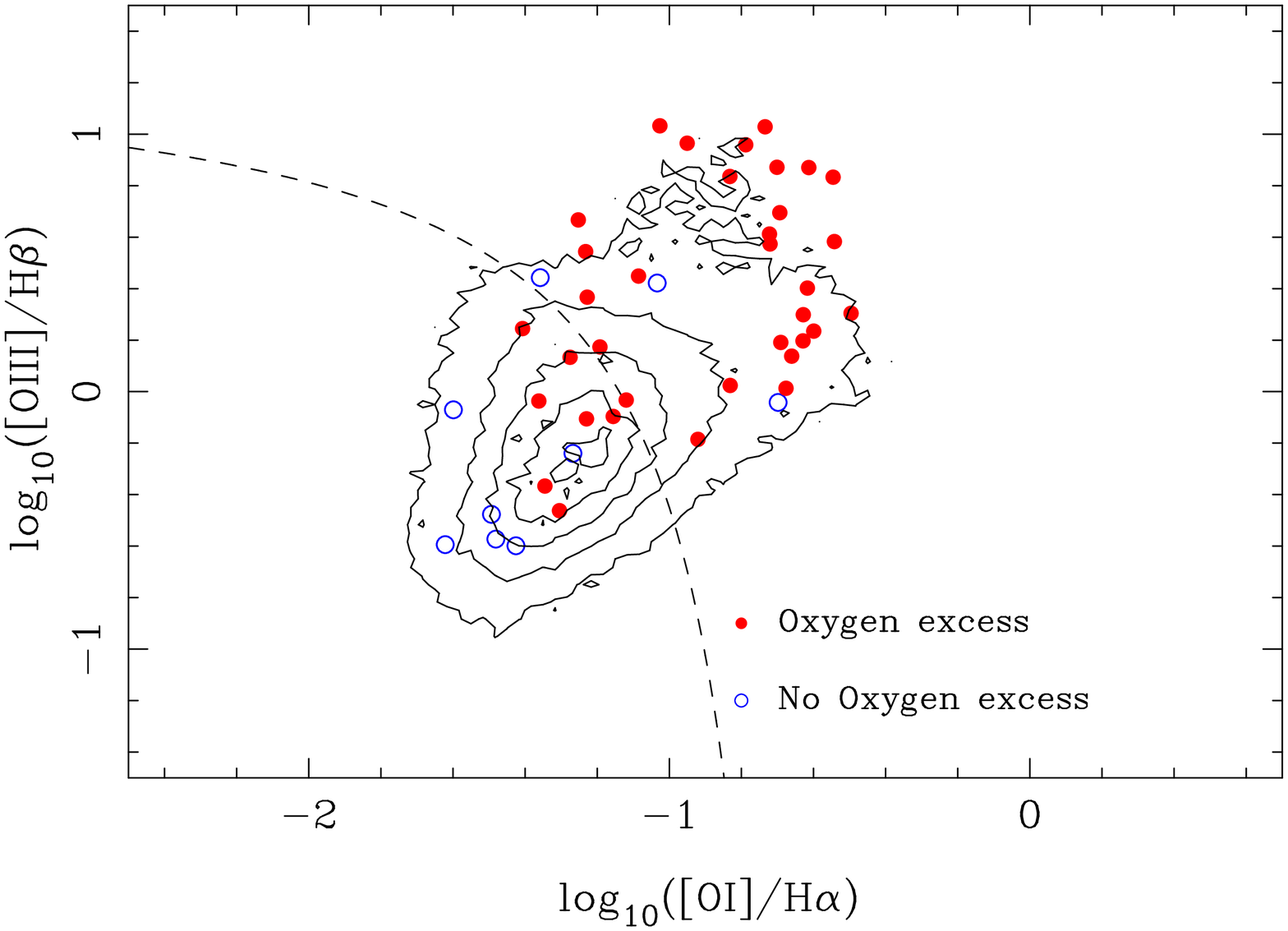}\hspace{0.5cm}
    \FigureFile(80mm,80mm){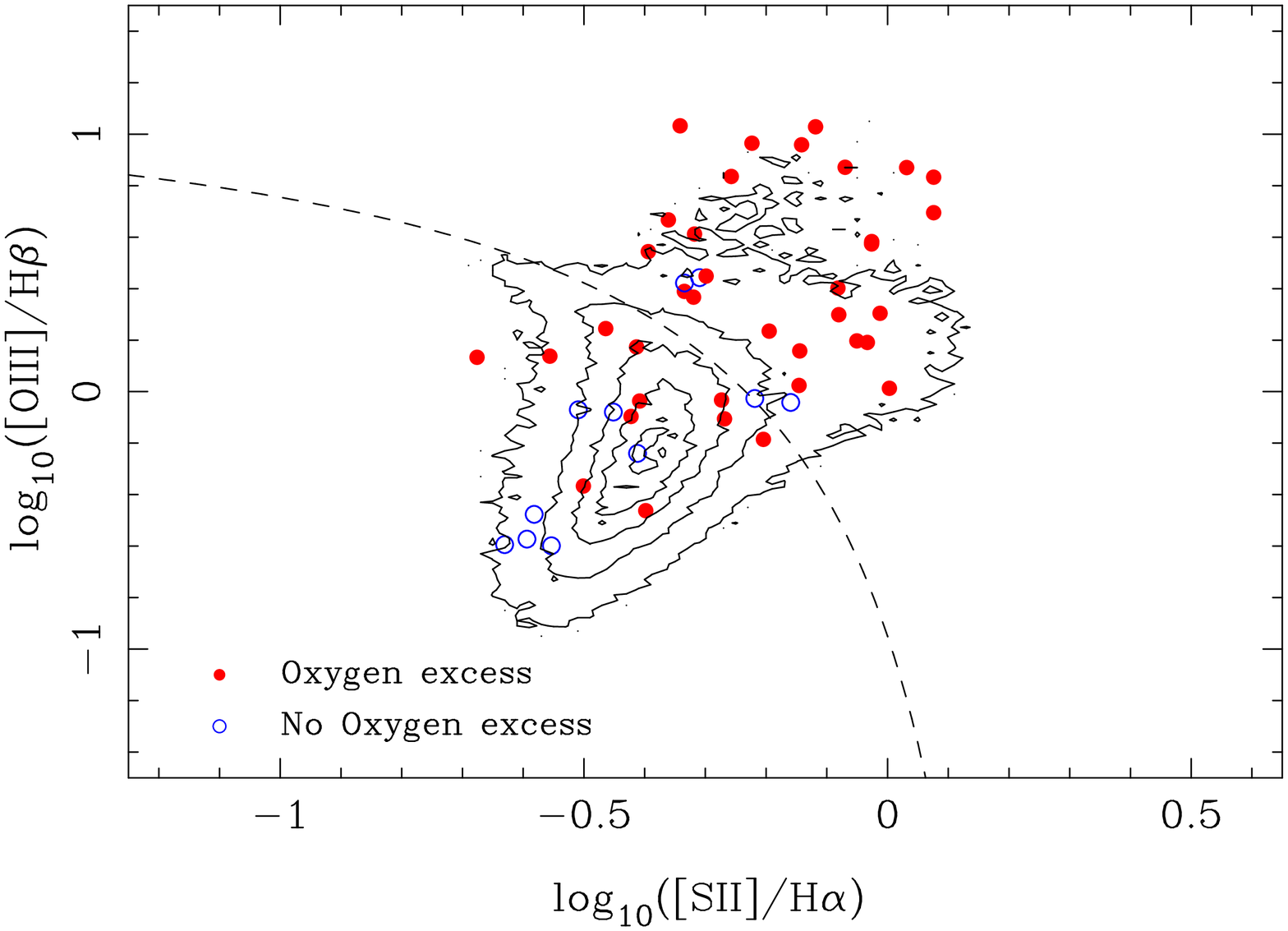}\\
  \end{center}
  \caption{
    As in Fig. \ref{fig:bpt_xray}, but for the two \citet{veilleux87} diagrams.
  }
  \label{fig:bpt2_xray}
\end{figure*}

\begin{figure*}
  \begin{center}
    \FigureFile(80mm,80mm){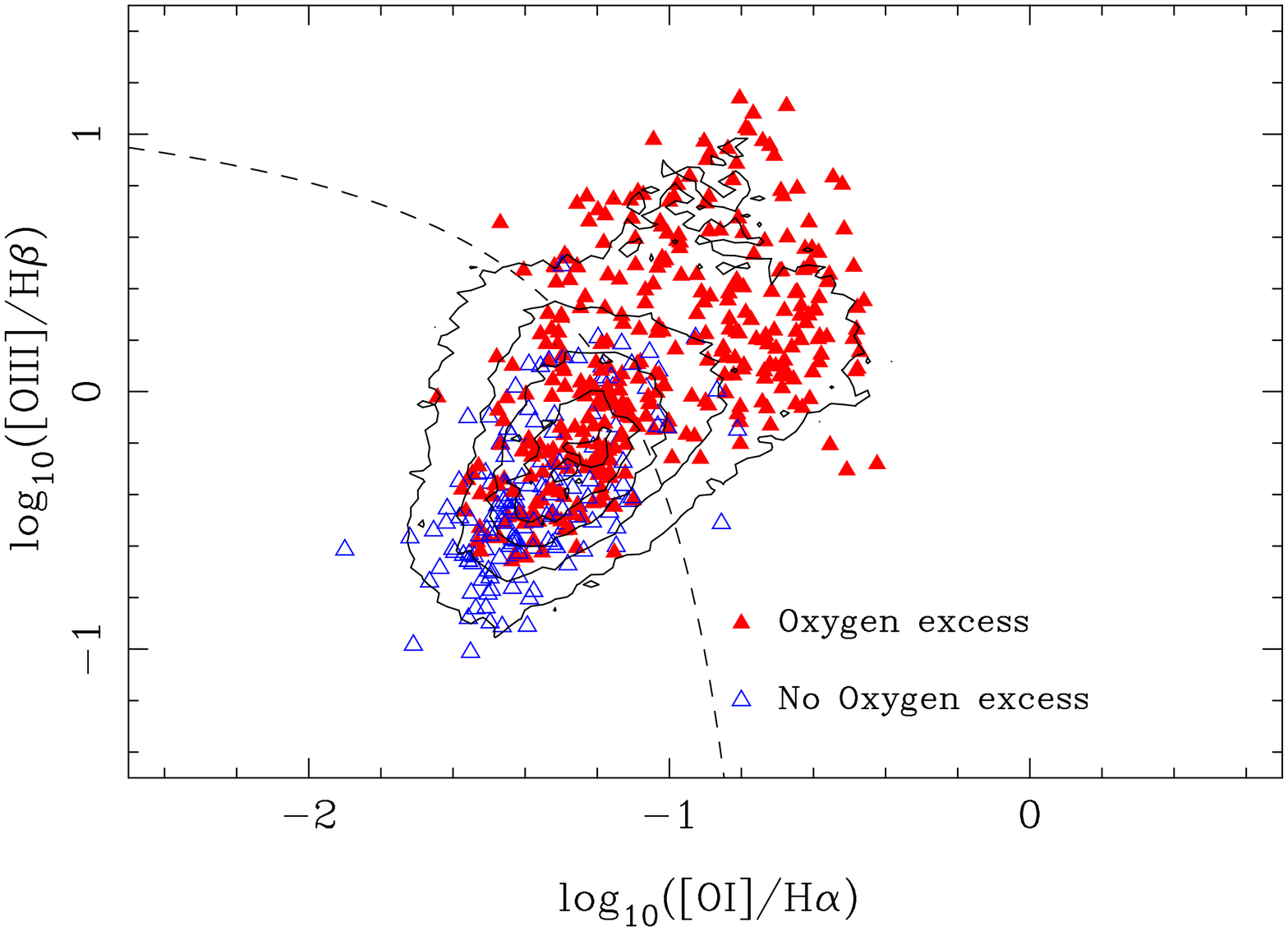}\hspace{0.5cm}
    \FigureFile(80mm,80mm){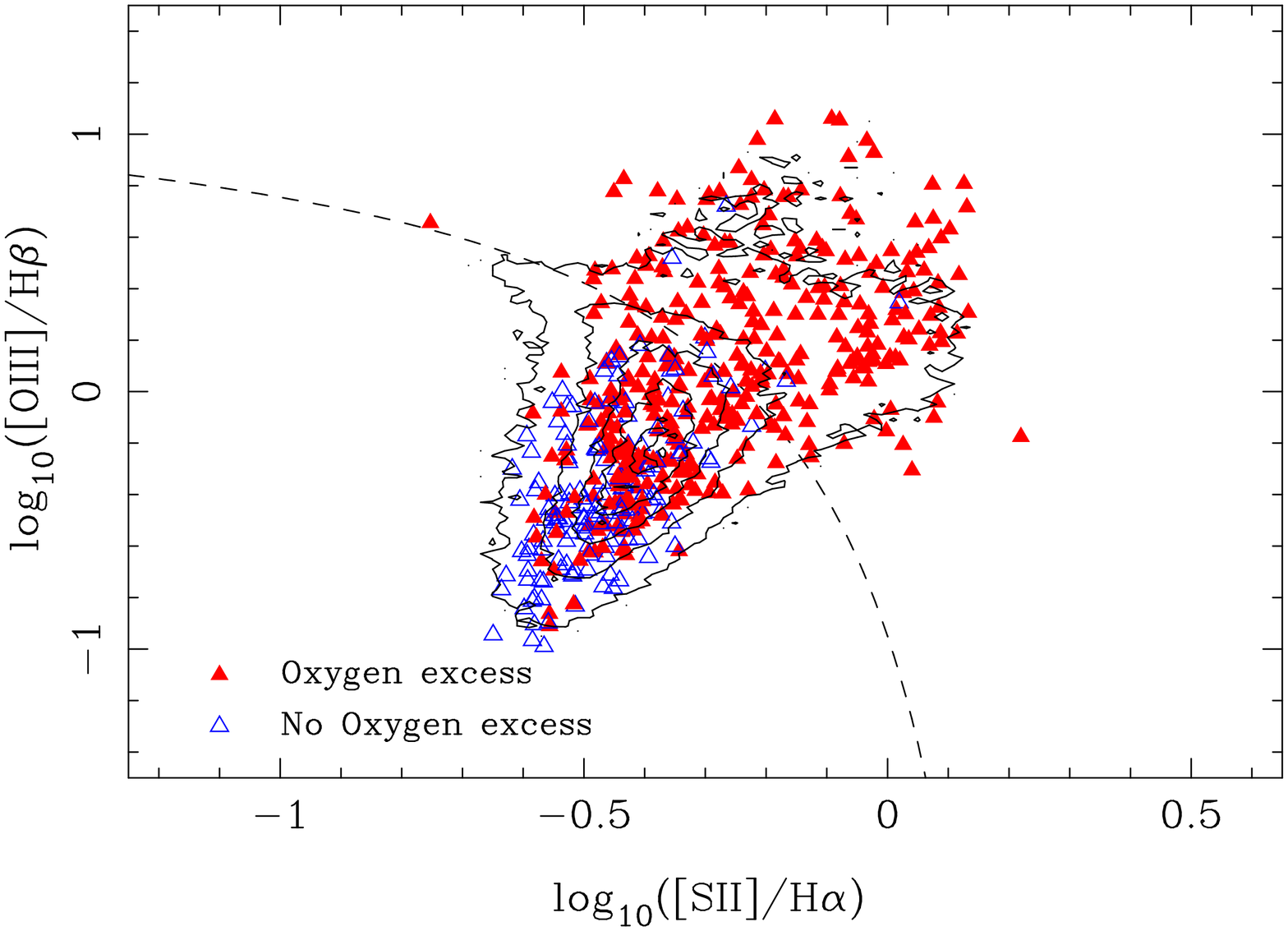}\\
  \end{center}
  \caption{
    As in Fig. \ref{fig:bpt_radio}, but for the two \citet{veilleux87} diagrams.
  }
  \label{fig:bpt2_radio}
\end{figure*}

The distribution of X-ray objects is shown in Fig. \ref{fig:bpt2_xray}.
Many of the sources are in the AGN region of the diagram, but 
a fraction of AGNs is scattered to the star forming region.
This may appear in contrast to the BPT diagram in Fig. \ref{fig:bpt_xray},
where we have observed that most X-ray sources are in the AGN region
of the diagram.
Radio sources are shown in Fig. \ref{fig:bpt2_radio}.
The radio objects spread over the diagram, but 
objects that do not show any Oxygen excess tend to lie
at the bottom-left tip of the star forming sequence.
These objects are likely actively forming stars and the radio
emission is due to star formation.

Overall, the \citet{veilleux87} diagrams show consistent results with BPT,
although their sensitivity to intermediate classes is limited.
As discussed in the main body of the paper, most of the Oxygen-excess
objects are in the O+Bn class and the figures presented in this appendix
show that O+Bn objects are clearly in the AGN region of the diagrams.
This adds further evidence that most of the Oxygen-excess method
works fairly well in identifying AGNs.


\section{Host galaxy properties of O-B+ objects}

The Oxygen-excess method misses a fraction of BPT AGNs (O-B+)
due possibly to active underlying star formation.  It could also be due to
strong featureless continuum from AGN, which affects
our SFR estimates.  We cannot easily distinguish these two
possibilities as discussed in the main body of the paper.
We thus do not try to characterize their AGN activities and
host galaxy SFRs.  Instead, we quantify their stellar mass,
color, and morphological types of the hosts and show
that these missing AGNs do not change our conclusions in Paper-II.

\begin{figure}
  \begin{center}
    \FigureFile(80mm,80mm){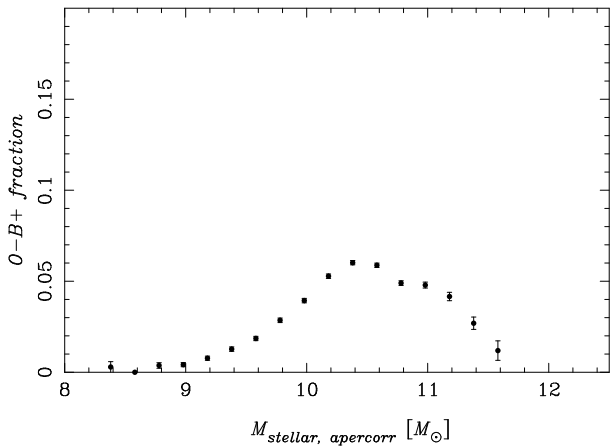}
  \end{center}
  \caption{
    Fraction of O-B+ objects as a function of stellar mass.
  }
  \label{fig:agn_comp_ob}
\end{figure}

Fig. \ref{fig:agn_comp_ob} shows the O-B+ fraction as a function
of stellar mass of the host galaxies.  The O-B+ objects typically
have $10^{10-11}\rm\ M_\odot$ and the fraction is low at the
most massive and least massive ends.  As we show in Paper-II,
the fraction of the Oxygen-excess objects is nearly 60\% at
$10^{11}\rm\ M_\odot$.  The fraction of the missing AGNs 
is an order of magnitude lower and they do not affect our results
in any significant way.

\begin{figure*}
  \begin{center}
    \FigureFile(80mm,80mm){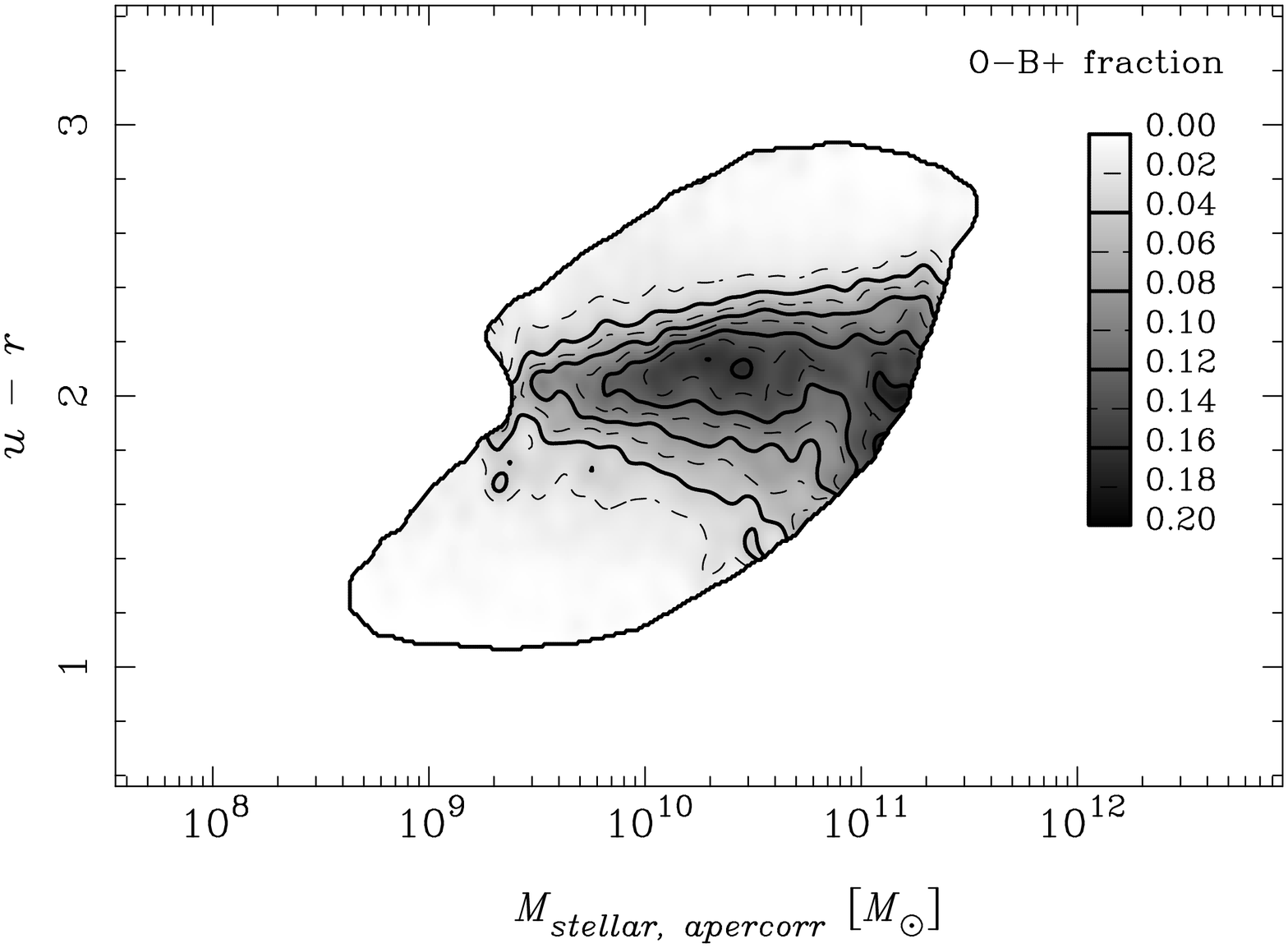}\hspace{0.5cm}
    \FigureFile(80mm,80mm){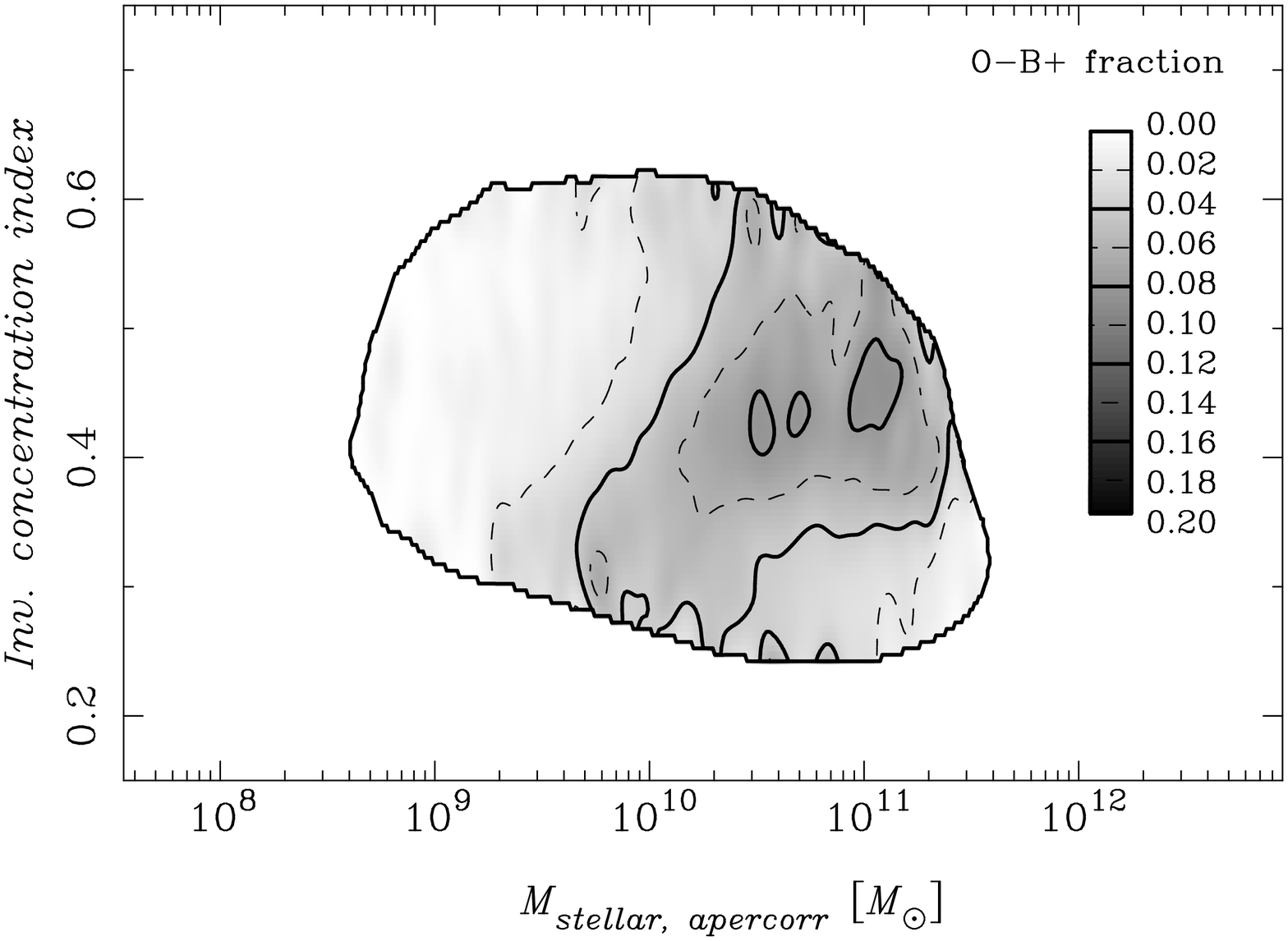}\\
  \end{center}
  \caption{
    O-B+ fraction as functions of rest-frame $u-r$ color and stellar mass (left),
    and inverse concentration index and stellar mass (right).
    Note that the contours are drawn at lower levels than those in Paper-II
    to clarify the distribution of the O-B+ objects.
  }
  \label{fig:agn_comp90_ob}
\end{figure*}

We plot the color and morphology distribution in Fig. \ref{fig:agn_comp90_ob}.
As detailed in Section 4 of Paper-II, the color is $k$-corrected
rest-frame $u-r$ color and the morphology is characterized
with the inverse concentration index \citep{shimasaku01,strateva01}
measured in the $z$-band.  The O-B+ objects tend to have
intermediate color and morphological types, which is similar to
the overall properties of the BPT AGNs (see paper-II).
They avoid the red sequence, which shows that they are undergoing star formation.

Overall, properties of the O-B+ objects are similar to
the BPT AGNs in general.
We have confirmed that our conclusions in Paper-II 
remain unchanged if we include these missing population in the analysis.

\end{document}